%% file: lmcs-long.tex
\documentclass{lmcs} %%% last changed 2014-08-20
\pdfoutput=1

\usepackage{lastpage}
\lmcsdoi{22}{2}{17}
\lmcsheading{}{\pageref{LastPage}}{}{}%
{Oct.~15,~2024}{May~22,~2026}{}

\usepackage{hyperref}

\usepackage[T1]{fontenc}
\usepackage[utf8]{inputenc}
\usepackage{tikz}
\usepackage{quiver}
\usepackage{adjustbox}

\input{packages.tex}

\input{chairmacros.tex}

\usepackage{hyperref}
\usepackage{color}

\csname endofdump\endcsname
\endofdump

\newcommand{\Qed}{}

\def\orcidID#1{\smash{\href{http://orcid.org/#1}{\protect\raisebox{1pt}{\protect\includegraphics{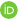}}}}}

\input{notation.tex}

\begin{document}

% If the title is longer than 55 characters, then specify a shorter running title as the optional argument to \title. The running title should be roughyl at most 55 characters:
%\title[Instructions]{Instructions for Authors\texorpdfstring{\\}{. }How to prepare papers
%  for LMCS using \texorpdfstring{\MakeLowercase{\texttt{lmcs.cls}}}{lmcs.cls}\rsuper*\\Version of
%  2022-04-01}
%\titlecomment{{\lsuper*}OPTIONAL comment concerning the title, \eg,
%  if a variant or an extended abstract of the paper has appeared elsewhere.}
%\thanks{thanks, optional.}	%optional

\title{A Resolution-Based Interactive Proof System for UNSAT}
%\titlerunning{A Resolution-Based Interactive Proof System for UNSAT}

% affiliations are numbered automatically with a, b, c (see below)
% use the optional argument to indicate the affiliation(s) of each author
% omit the argument if there is only one author, or only one affiliation
\author[P.~Czerner]{Philipp Czerner\lmcsorcid{0000-0002-1786-9592}}
\author[J.~Esparza]{Javier Esparza\lmcsorcid{0000-0001-9862-4919}}
\author[V.~Krasotin]{Valentin Krasotin\lmcsorcid{0009-0002-2129-2754}}
\author[A. Krauss]{Adrian Krauss\lmcsorcid{0009-0003-2693-0740}}

% affiliation 1 (automatically numbered a)
\address{Technical University of Munich}	%optional
% write emails for all authors having that affiliation
\email{\{czerner, esparza, krasotin\}@cit.tum.de, adrian.krauss@tum.de}  %optional

% affiliation 2 (automatically numbered b)
%\address{University 2, address2}	%optional
%\email{name2@email2}  %optional

%% etc.

%% required for running head on odd and even pages, use suitable
%% abbreviations in case of long titles and many authors:

%%%%%%%%%%%%%%%%%%%%%%%%%%%%%%%%%%%%%%%%%%%%%%%%%%%%%%%%%%%%%%%%%%%%%%%%%%%

%% the abstract has to PRECEDE the command \maketitle:
%% be sure not to issue the \maketitle command twice!

\begin{abstract}
Modern SAT or QBF solvers are expected to produce correctness certificates. However, UNSAT certificates have worst-case exponential size (unless $\NP=\coNP$), and at recent SAT competitions the largest certificates of unsatisfiability are reaching terabyte size. This puts limits to the development of SAT-solving services in which a client with limited computational power sends a formula to a solver running on a powerful server, which returns a certificate to be checked by the client.

Recently, Couillard, Czerner, Esparza,
and Majumdar have suggested to replace certificates with interactive proof systems based on the $\textsf{IP}=\textsf{PSPACE}$ theorem. They have presented an interactive protocol between a prover and a verifier
for an extension of QBF. The overall running time of the protocol is linear in the time
needed by a standard BDD-based algorithm, and the time invested by the verifier is polynomial in the size of the formula. (So, in particular, the verifier never has to read or process exponentially long certificates).
We call such an interactive protocol \emph{competitive} with the BDD algorithm for solving QBF.

While BDD algorithms are state-of-the-art for certain classes of QBF instances, no modern (UN)SAT solver is based on BDDs. For this reason, we initiate the study of interactive certification for more practical SAT algorithms. In particular, we address the question whether interactive protocols can be competitive with some variant of resolution. We present two contributions. First, we prove a theorem that reduces the problem of finding competitive interactive protocols to finding an \emph{arithmetisation} of formulas satisfying certain commutativity properties. (Arithmetisation is the fundamental technique underlying the $\textsf{IP}=\textsf{PSPACE}$ theorem.) Then, we apply the theorem to give the first interactive protocol for the Davis-Putnam resolution procedure. We also report on an implementation and give some experimental results.

Parts of this work appeared previously as~\cite{CzernerEK24}.
\end{abstract}

\maketitle

%% start the paper here:

\section{Introduction}
Automated reasoning tools should provide evidence of their correct behaviour.
A substantial amount of research has gone into proof-producing automated reasoning tools~\cite{Heule21,Necula97,Namjoshi01,HenzingerJMNSW02,BarbosaRKLNNOPV22}. 
These works define a notion of ``correctness certificate'' and adapt the reasoning engine to produce independently checkable certificates. For example, SAT solvers produce either a satisfying assignment or a proof of unsatisfiability in some proof system, e.g.\ resolution (see~\cite{Heule21} for a survey). We call them SAT and UNSAT certificates, respectively.

There is a fundamental asymmetry between SAT and UNSAT certificates: while SAT certificates can be checked in polynomial (even linear) time in the size of the formula, this is very unlikely to be the case for UNSAT certificates, because it implies $\textsf{NP}=\textsf{coNP}$. In particular, all UNSAT certificates produced by current tools (like a DRAT certificate or a resolution proof) may have exponential size in the size of the formula, and UNSAT certificates produced by current tools may have hundreds of GiB or even, in extreme cases, hundreds of TiB~\cite{HeuleKM16}. Despite much progress on reducing the size of proofs and improving the efficiency of checking proofs (see e.g.~\cite{HeuleHKW17,Heule21}), this fundamental asymmetry puts limits to the development of SAT-solving services in which clients with limited computational power (say, a standard laptop) send formulas to a solver running on a very powerful server, and get back a SAT or UNSAT certificate. 
We are interested in such a SAT-solving scenario, where the costs of solving and certificate checking are assumed by different parties, and reducing the latter is important, even if it increases the former. Notice that this stands in contrast to the cost model of a SAT-solving tool that solves the formula, produces a certificate, and checks it, where one is only interested in minimising the total time.

The $\textsf{IP}=\textsf{PSPACE}$ theorem, proved in 1992 by Shamir~\cite{Shamir92}, presents a possible fundamental solution to this problem: \emph{interactive proofs}\footnote{In our context it would be more adequate to speak of interactive certification, but we use the standard terminology.}. A language is in $\textsf{IP}$ if there exists a sound and complete \emph{interactive proof protocol} between two agents, Prover and Verifier, that Verifier can execute in randomised polynomial time~\cite{GoldwasserMicaliRackoff,Babai,LundFKN92,AB09}.  Completeness means that, for any input in the language, an \emph{honest prover} that truthfully follows the protocol will convince Verifier to accept the input. Soundness means that, for any input not in the language, Verifier will reject it with high probability, no matter how Prover behaves. ``Conventional'' certification is the special case of interactive proof in which Prover sends Verifier only one message, the certificate, and Verifier is convinced with probability 1. The $\textsf{IP}=\textsf{PSPACE}$ theorem implies the existence of interactive proof protocols for UNSAT in which Verifier only invests polynomial time \emph{in the size of the formula}. 
This is not the case for conventional certification protocols where Prover produces a certificate, like a resolution proof or a DRAT certificate. Indeed, while these certificates can be checked in linear time \emph{in the size of the certificate}, they have worst-case exponential size in the size of the formula, and so Verifier needs exponential time in the worst case.

Despite its theoretical promise, the automated reasoning community has not yet developed tools for UNSAT or QBF with interactive proof protocols.  In a recent paper, Couillard, Czerner, Esparza, and Majumdar venture a possible explanation~\cite{CouillardCEM23}. They observe that the interactive certification protocols described in the literature have been designed to prove asymptotic complexity results, for which one works with a cost model that only counts the resources used by Verifier, that is, a model in which the steps of Prover have zero cost. Intuitively, in this approach the goal is to make Verifier faster, \emph{regardless} of the cost for Prover. These leads to very inefficient Provers that typically construct the complete truth table of the formula, making them incompatible with automated reasoning tools. Indeed, after efficiently deciding that a given formula is UNSAT using a state-of-the-art algorithm, Prover would have to compute the truth table of the formula. The runtime of Prover would be \emph{best-case} exponential in the number of variables, making the tool useless in practice.

Couillard \textit{et al.}\ propose to search for interactive proof protocols where Prover uses algorithms that, while worst-case exponential, like any known algorithm for QBF, are more efficient in practice than brute force. In~\cite{CouillardCEM23} they consider the standard BDD-based algorithm for QBF and design an interactive protocol based on it.

In this paper, we initiate the study of interactive certification for UNSAT for Provers based on more efficient SAT-solving algorithms than pure brute-force. For this, given an algorithm $\Algorithm$ and an interactive protocol $P$, both for UNSAT, we say that $P$ is \emph{competitive} for $\Algorithm$ if the ratio between the runtime of Prover in $P$ and the runtime of $\Algorithm$ on inputs of length $n$ is bounded by a polynomial in $n$. So, loosely speaking, if $P$ is competitive for $\Algorithm$, then one can add interactive verification to $\Algorithm$ with only polynomial overhead. For example, Provers based on computing the truth table are not competitive with respect to any (reasonable variant of) resolution: Indeed, it is easy to construct a family of formulas with an arbitrary number of variables such that resolution takes polynomial time in the size of the formula, while constructing the truth table takes exponential time. This raises the question whether it is possible to design interactive protocols that are competitive for algorithms more efficient than brute-force methods.

We present the first interactive protocol competitive for (a simplified version of) the well-known Davis-Putnam procedure---see e.g.\ Section~2.9 of~\cite{Harrison09}. Our version fixes a total order on variables, resolves exhaustively with respect to the next variable, say $x$, and then ``locks'' all clauses containing $x$ or $\neg x$, ensuring that they are never resolved again w.r.t.\ any variable. For this, we first present the gist of our idea in general terms. We introduce a technique that, given an algorithm $\Algorithm$ for UNSAT satisfying certain conditions, constructs an interactive protocol that is competitive with respect to $\Algorithm$.  Let us be more precise. We consider algorithms for UNSAT that, given a formula $\phii{0}$, construct a sequence $\phii{0}, \phii{1}, \ldots, \phii{k}$ of formulas such that $\phii{i}$ is equisatisfiable to $\phii{i+1}$, and there is a polynomial algorithm that decides if $\phii{k}$ is unsatisfiable.  Our interactive protocols are based on the well-known idea of encoding the formulas in this sequence as polynomials over a finite field in such a way that the truth value of the formula for a given assignment is determined by the value of the polynomial on that assignment. The encoding procedure is called  \emph{arithmetisation}, and has been extensively used for designing both SAT-solving algorithms (see e.g.~\cite{BussN21} on algebraic proof systems) and interactive proof protocols~\cite{GoldwasserMicaliRackoff,Babai,LundFKN92,AB09}. We introduce the notion of an arithmetisation \emph{compatible} with a given algorithm. Loosely speaking, compatibility means that for each step $\phii{i} \mapsto \phii{i+1}$, there is an operation on polynomials mimicking the operation on formulas that transforms $\phii{i}$ into $\phii{i+1}$. We prove that the problem of finding a competitive interactive protocol for a given algorithm $\Algorithm$ for UNSAT reduces to finding an arithmetisation compatible with $\Algorithm$. 
We then show that, while standard arithmetisations are not compatible with the Davis-Putnam algorithm, a non-standard arithmetisation is. In our opinion, this is the main insight of our paper: in order to find interactive protocols for algorithms for UNSAT, one can profit from non-standard arithmetisations. 

The Davis-Putnam procedure was proposed in the 1960s, and constitutes a first link in a chain of increasingly efficient SAT-solving procedures that includes different flavours of resolution, DPLL and, ultimately, conflict-driven clause learning (CDCL). Modern techniques for solving UNSAT are several orders of magnitude faster than Davis-Putnam, and so our results do not yet have any practical implication for SAT solving: in order to achieve polynomial time for Verifier, we would need to use an algorithm for Prover that, while much better than brute force, is still too inefficient compared to the current state of the art. Whether practical interactive protocols can be obtained for other links of the chain remains open. On the one hand, our technique does not extend  to them, and so, besides non-standard arithmetisations, new ideas are required. On the other hand, complexity theory offers several candidates for such ideas: besides the \texttt{IP} model used in our paper, one may also consider the multi-prover model \texttt{MIP}, as well as probabilistically checkable proofs.

In the last section of the paper, we report on an implementation of our interactive protocol. We conduct some experiments comparing the time and memory consumption of Prover and Verifier for Davis-Putnam in both the conventional and interactive certification settings. In the conventional setting, Prover computes a Davis-Putnam resolution proof and sends it to Verifier, who checks its correctness. In the interactive setting, Prover still computes the resolution proof but, instead of sending it to Verifier, engages with Verifier in our competitive interactive protocol.  Roughly speaking, our results indicate that Prover is slower in the interactive setting by a nearly constant factor, while Verifier is faster in the interactive setting by several orders of magnitude.

The paper is structured as follows. Section~\ref{sec:prelims} contains preliminaries. Section~\ref{sec:ips} presents interactive proof systems and defines interactive proof systems competitive with a given algorithm. Section~\ref{sec:fullres} defines our version of the Davis-Putnam procedure. Section~\ref{sec:abstractprotocol} introduces arithmetisations, and defines arithmetisations compatible with a given algorithm. Section~\ref{sec:competitive} presents an interactive proof system for Davis-Putnam. Section~\ref{sec:impl} reports on our implementation and experiments. Section~\ref{sec:conclusions} contains conclusions.

\section{Preliminaries}
\label{sec:prelims}
\paragraph{Multisets.} A multiset over a set $S$ is a mapping $m \colon S \rightarrow \mathbb{N}$. We also write multisets using set notation, for example we write $\{x,x,y\}$ or $\{2\cdot x, y\}$. Given two multisets $m_1$ and $m_2$, we define  $m_1 \oplus m_2$ as the multiset given by $(m_1 \oplus m_2)(s) = m_1(s) + m_2(s)$ for every $s \in S$, and $m_1 \ominus m_2$ as the multiset given by $(m_1 \ominus m_2)(s) = \max\{ 0, m_1(s) - m_2(s)\}$ for every $s \in S$.

\paragraph{Formulas, CNF, and resolution.}  A Boolean \emph{variable} has the form $x_i$ where $i = 1,2,3,...$. Boolean \emph{formulas} are defined inductively: $\True$, $\False$ and variables are formulas; if $\varphi$ and $\psi$ are formulas, then so are $\neg \varphi$, $\varphi \vee \psi$, and $\varphi \wedge \psi$. A \emph{literal} is a variable or the negation of a variable. A formula $\varphi$ is in \emph{conjunctive normal form (CNF)} if it is a conjunction of disjunctions of literals. We represent a formula in CNF as a \emph{multiset} of \emph{clauses} where a clause is a \emph{multiset} of literals. For example, the formula $(x \vee x \vee x \vee \neg y) \wedge z \wedge z $ is represented by the multiset $\{ \{3x,\neg y\}, 2 \{z\} \}$.

\begin{rem}
Usually CNF formulas are represented as \emph{sets} of clauses, which are defined as \emph{sets} of literals. 
Algorithms that manipulate CNF formulas using the set representation are assumed to silently remove duplicate formulas or duplicate literals. In this paper, due to the requirements of interactive protocols, we need to make these steps explicit. In particular, we use multiset notation for clauses. For example, $C(x)$ denotes the number of occurrences of $x$ in $C$.
\end{rem}

We assume in the paper that formulas are in CNF. Abusing language, we use $\varphi$ to denote both a (CNF) formula and its multiset representation.

%The set representation of a CNF formula retains associativity, commutativity, and idempotence of $\vee$ and $\wedge$, as elements in a set do not have an order, and a set containing an element multiple times is equivalent to the set containing it once.

%\newcommand{\res}[1]{\textit{Res}_{#1}}
\paragraph{Resolution.} Resolution is a proof system for CNF formulas. Given a variable $x$, a clause $C$ containing exactly one occurrence of $x$ and a clause $C'$ containing exactly one occurrence of $\neg x$, the \emph{resolvent} of $C$ and $C'$ with respect to $x$ is the clause $\res{x}(C, C'):=(C \ominus \{x\}) \oplus (C' \ominus \{\neg x\})$.

For example, $\res{x}(\{x,\neg y, z\},\{\neg x, \neg w\}) = \{\neg y, z, \neg w\}$. It is easy to see that $C \wedge C'$ and $\res{x}(C,C')$ are equisatisfiable. A \emph{resolution refutation} for a formula $\varphi$ is a sequence of clauses ending in the empty clause whose elements are either clauses of $\varphi$ or resolvents of two previous clauses in the sequence. It is well known that $\varphi$ is unsatisfiable if{}f there exists a resolution refutation for it. There exist families of formulas, like the pigeonhole formulas, for which the length of the shortest resolution refutation grows exponentially in the size of the formula, see e.g.~\cite{Haken85,BussT88}. 

\paragraph{Polynomials.}
Interactive protocols make extensive use of polynomials over a finite field $\F$. 
%We assume the reader is familiar with polynomials. 
Let $X$ be a finite set of variables. We use $x, y, z, \ldots$ for variables and $p, p_1, p_2, \ldots$ for polynomials.
%When we write a polynomial explicitly, we write it in brackets, e.g.\ $[3xy-z^2]$.
%We write $\One$ and $\Zero$ for the polynomials $[1]$ and $[0]$, respectively.
%We write $\Polynom{X}$ for the set of all polynomials over $X$.
We use the following operations on polynomials:
\begin{itemize}
\item \emph{Sum, difference, and product}, denoted $p_1+p_2$, $p_1-p_2$, $p_1 \cdot p_2$, and defined as usual. For example, $(3xy-z^2)+(z^2+yz)=3xy+yz$ and $(x+y)\cdot(x-y)=x^2-y^2$.
\item \emph{Partial evaluation}. Denoted $\pev{x \Gets a} \, p$, it returns the result of setting the variable $x$ to the field element $a$ in the polynomial $p$, e.g.\ $\pev{x \Gets  5}(3xy-z^2)=15y-z^2$.
\end{itemize}

A \emph{(partial) assignment} is a (partial) mapping $\sigma: X \rightarrow \F$. We write $\Pev\sigma \, p$ instead of $\pev{x_1:=\sigma(x_1)}...\pev{x_k:=\sigma(x_k)} \, p$, where $x_1,...,x_k$ are the variables for which $\sigma$ is defined. Additionally, we call a (partial) assignment $\sigma$ \emph{binary} if $\sigma(x)\in\{0,1\}$ for each $x\in X$.

The following lemma is at the heart of all interactive proof protocols. Intuitively, it states that if two polynomials are different, then they are different for almost every input. Therefore, by picking an input at random, one can check polynomial equality with high probabillity.

\begin{lem}[Schwartz-Zippel Lemma]
\label{lem:sz}
Let $p_1, p_2$ be distinct univariate polynomials over $\F$ of degree at most $d \geq 0$. Let $r$ be selected uniformly at random from $\F$. The probability that $p_1(r) = p_2(r)$ holds is at most $d /\Abs{\F}$.
\end{lem}
\begin{proof}
Since $p_1\ne p_2$, the polynomial $p := p_1 - p_2$ is not the zero polynomial and has degree at most $d$. Therefore $p$ has at most $d$ zeros, and so the probability of $p(r)=0$ is at most $d / \Abs{\F}$.\Qed
\end{proof}

\section{Interactive Proof Systems}
\label{sec:ips}
%\parag{The Class $\IP$}
An \emph{interactive protocol} is a sequence of interactions between two parties: \emph{Prover} and \emph{Verifier}. Prover has unbounded computational power, whereas Verifier is a randomised, polynomial-time algorithm. Initially, the parties share an input $x$ that Prover claims belongs to a given language $L$ (e.g. UNSAT). The parties alternate in sending messages to each other according to a protocol. Intuitively, Verifier repeatedly asks Prover to send informations. At the end of the protocol, Verifier accepts or rejects the input. 

Formally, let $V,P$ denote (randomised) online algorithms, i.e.\ given a sequence of inputs $m_1,m_2,...\in\{0,1\}^*$ they compute a sequence of outputs, e.g.\ $V(m_1), V(m_1,m_2), ...$. We say that $(m_1,...,m_{2k})$ is a \emph{$k$-round interaction}, with $m_1,...,m_{2k}\in\{0,1\}^*$, if $m_{i+1}=V(m_1,...,m_i)$ for odd $i$ and $m_{i+1}=P(m_1,...,m_i)$ for even $i$.

The \emph{output} $\Out_{V,P,k}(x)$ is $m_{2k}$, where $(m_1,...,m_{2k})$ is a $k$-round interaction with $m_1=x$.
We also define the \emph{Verifier-time} $\Vtime_{V,P,k}(x)$ as the expected time it takes $V$ to compute $m_2,m_4,...,m_{2k}$ for any $k$-round interaction $(m_1,...,m_{2k})$ with $m_1=x$. We define the \emph{Prover-time} $\Ptime_{V,P,k}(x)$ analogously.

Let $L$ be a language and $p:\N\rightarrow\N$ a polynomial. A tuple $(V,P_H,p)$ is an \emph{interactive protocol for $L$} if for each $x\in\{0,1\}^*$ of length $n$ we have $\Vtime_{V,P_H,p(n)}(x)\in\O(\Poly n)$ and:
\begin{enumerate}
\item (\emph{Completeness}) $x\in L$ implies $\Out_{V,P_H,p(n)}(x)=1$ with probability 1, and
\item (\emph{Soundness}) $x\notin L$ implies that for all $P$ we have $\Out_{V,P,p(n)}(x)=1$ with probability at most $2^{-n}$.
\end{enumerate}
The completeness property ensures that if the input belongs to the language $L$, then there is an “honest” Prover $P_H$ who can always convince Verifier that indeed $x\in L$. If the input does not belong to the language, then the soundness property ensures that Verifier rejects the input with high probability no matter how a (dishonest) Prover tries to convince it. 

$\IP$ is the class of languages for which there exists an interactive protocol. It is known that $\IP = \PSPACE$~\cite{LundFKN92,Shamir92}, that is, every language in $\PSPACE$ has a polynomial-round interactive protocol.
The proof exhibits an interactive protocol for the language QBF of true quantified boolean formulas; in particular, 
the honest Prover is a polynomial-space, exponential-time algorithm.

\subsection{Competitive Interactive Protocols}
In an interactive protocol  there are no restrictions on the running time of Prover. The existence of an interactive protocol for some \textsf{coNP}-complete problem in which Prover runs in polynomial time would imply e.g.\ $\NP\subseteq\textsf{BPP}$. Since this is widely believed to be false, Provers are allowed to run in exponential time, as in the proofs of~\cite{LundFKN92,Shamir92}. However, while all known approaches for UNSAT use exponential time in the worst case, some perform much better in practice than others. For example, the Provers of~\cite{LundFKN92,Shamir92} run in exponential time \emph{in the best case}. This motivates our next definition: instead of stating that Prover must always be efficient, we say that it must have a bounded overhead \emph{compared} to some given algorithm $\Algorithm$.

Formally, let $L\subseteq \{0,1\}^*$ be a language, let $\Algorithm$ be an algorithm for $L$, and let $(V,P_H,p)$ be an interactive protocol for $L$. We say that $(V,P_H,p)$ is \emph{competitive with $\Algorithm$} if for every input  $x\in\{0,1\}^*$ of length $n$ we have $\Ptime_{V,P_H,p(n)}(x)\in\O(\Poly(n)T(x))$, where $T(x)$ is the time it takes $\Algorithm$ to run on input $x$.

Recently, Couillard, Czerner, Esparza and Majumdar~\cite{CouillardCEM23} have constructed an interactive protocol for QBF that is competitive with \BDDSolver, the straightforward BDD-based algorithm that constructs a BDD for the satisfying assignments of each subformula, starting at the leaves of the syntax tree and progressively moving up. %In their construction, $p(n)$ is linear or even constant, depending on the implementation.
In this paper, we will investigate UNSAT and give an interactive protocol that is competitive with \FullResolution, a decision procedure for UNSAT based on a restricted version of resolution.

\section{The Davis-Putnam Resolution Procedure}
\label{sec:fullres}

We introduce the variant of the Davis-Putnam resolution procedure ~\cite{DavisP60,Harrison09} for which we later construct a competitive interactive protocol\footnote{In Harrison's book~\cite{Harrison09}, the Davis-Putnam procedure consists of three rules. The version in Definition~\ref{def:fullresstep} uses only Rule III, which is sometimes called the Davis-Putnam resolution procedure.  Unfortunately, at the time of writing this paper, the Wikipedia article for the Davis-Putnam algorithm uses a different terminology (even though it cites~\cite{Harrison09}): it calls the three-rule procedure the Davis-Putnam \emph{algorithm}, and the algorithm consisting only of Rule III the Davis-Putnam \emph{procedure}.}. Recall that in our setting, clauses are multisets, and given a clause $C$ and a literal $l$, $C(l)$ denotes the number of occurrences of $l$ in $C$.

\begin{defi}\label{def:fullresstep}
Let $x$ be a variable. \emph{Full $x$-resolution} is the procedure that takes as input a formula $\varphi$ satisfying $C(x) + C(\neg x) \leq 1$ for every clause $C$, and returns the formula $R_x(\varphi)$ computed as follows:
\begin{enumerate}
    \item For every pair $C_1,C_2$ of clauses of $\varphi$ such that $x \in C_1$ and $\neg x \in C_2$, add to $\varphi$ the resolvent w.r.t. $x$ of $C_1$ and $C_2$ (i.e.\ set $\varphi := \varphi \oplus \res{x}(C_1,C_2)$).
    \item Remove all clauses containing $x$ or $\neg x$.
\end{enumerate}
\emph{Full $x$-cleanup} is the procedure that takes as input a formula $\varphi$ satisfying $C(x) + C(\neg x) \leq 2$ for every clause $C$, and returns the formula $C_x(\varphi)$ computed as follows:
\begin{enumerate}
\item Remove from $\varphi$ all clauses containing both $x$ and $\neg x$.
\item Remove from each remaining clause all duplicates of $x$ or $\neg x$.
\end{enumerate}
The \emph{Davis-Putnam resolution procedure} is the algorithm for \text{UNSAT} that, given a total order $x_1 \prec x_2 \prec \cdots \prec x_n$ on the variables of an input formula $\varphi$, executes Algorithm~\ref{alg:fr}. The algorithm assumes that $\varphi$ is a set of sets of literals, that is, clauses contain no duplicate literals, and $\varphi$ contains no duplicate clauses. We let $\Box$ denote the empty clause.
\end{defi}
Observe that while the initial formula contains no duplicate clauses, the algorithm may create them, and they are not removed.

\begin{algorithm}[t]
\caption{\FullResolution($\varphi$)}
\label{alg:fr}
\begin{algorithmic}
\For{$i=1, \ldots, n$}
	\State $\varphi:=R_{x_i}(\varphi)$
	\For{$j=i+1, \ldots, n$}  \State $\varphi:=C_{x_j}(\varphi)$
	\EndFor
\EndFor
\If{$\Box \in \varphi$} \State return ``unsatisfiable''
\Else \State return ``satisfiable''
\EndIf
\end{algorithmic}
\end{algorithm}

\begin{exa}
Table~\ref{tab:runningexample} shows on the left a run of \FullResolution\ on a formula $\varphi$ with three variables and six clauses.
The right column is explained  in Section~\ref{subsec:arithtwo}.
\begin{table}[t]
\caption{Run of \FullResolution\ on an input $\varphi$, and arithmetisation of the intermediate formulas.}
\label{tab:runningexample}
\smallskip
\centering
%\begin{center}
\resizebox{\textwidth}{!}{%
\begin{tabular}{c@{\hskip 2mm}l@{\hskip 5mm}l}
\toprule
Step & \multicolumn{1}{c}{Formula} & \multicolumn{1}{c}{Arithmetisation}\\
\midrule
Inp.
& 
$\begin{array}[t]{rcl}
\varphi & = & \{\{x,y\}, \{x, \neg y, \neg z\}, \{\neg x, \neg z\}, \\
& & \{\neg x, \neg y, \neg z\}, \{y,z\}, \{\neg y, z\}\}
\end{array}$
&
$\begin{array}[t]{rcl}
\ArithTwo(\varphi)   & = & (1-x)(1-y) + (1-x)y^3z^3 + x^3z^3\\
& & + x^3y^3z^3 + (1-y)(1-z) + y^3(1-z)
\end{array}$  
\\[0.7cm]
$R_x$ 
&   
$\begin{array}[t]{rcl}
\varphi_1 & = & \{\{y, \neg z\}, \{y, \neg y, \neg z\}, \{\neg y, \neg z, \neg z\} \\
& & \{\neg y, \neg z, \neg y, \neg z\}, \{y,z\}, \{\neg y, z\}\}
\end{array}$
&
$\begin{array}[t]{rcl}
\ArithTwo(\varphi_1) & = & (1-y)z^3 + (1-y)y^3z^3 + y^3z^6\\
& & + y^6z^6 + (1-y)(1-z) + y^3(1-z)
\end{array}$
\\[0.7cm]
$C_y$ 
& 
$\begin{array}[t]{rcl}
\varphi_2 & = & \{\{y, \neg z\}, 2 \cdot \{\neg y, \neg z, \neg z\}, \\
& & \{y,z\}, \{\neg y, z\}\} 
\end{array}$
&
$\begin{array}[t]{rcl}
\ArithTwo(\varphi_2)  & = & (1-y)z^3 + 2y^3z^6\\
& & + (1-y)(1-z) + y^3(1-z)
\end{array}$
\\[0.7cm]
$C_z$ 
&
$\begin{array}[t]{rcl}
\varphi_3 & = & \{\{y, \neg z\}, 2\cdot \{\neg y, \neg z\}  \\
& & \{y,z\}, \{\neg y, z\}\}
\end{array}$
&
$\begin{array}[t]{rcl}
\ArithTwo(\varphi_3)  & = & (1-y)z^3 + 2y^3z^3\\
& & +(1-y)(1-z) + y^3(1-z)
\end{array}$
\\[0.7cm]
$R_y$ 
& 
$\begin{array}[t]{rcl}
\varphi_4 & = & \{2\cdot \{\neg z, \neg z\}, 3\cdot \{\neg z, z\}, \{z, z\}\} \\
\end{array}$
&
$\begin{array}[t]{rcl}
\ArithTwo(\varphi_4)  & = & 2z^6 + 3z^3(1-z) + (1-z)^2
\end{array}$
\\[0.2cm]
$C_z$ 
& 
$\begin{array}[t]{rcl}
\varphi_5 & = & \{2 \cdot \{\neg z\}, \{z\}\} 
\end{array}$
&
$\begin{array}[t]{rcl}
\ArithTwo(\varphi_5)  & = & 2z^3 + (1-z) 
\end{array}$
\\[0.2cm]
$R_z$
&
$\begin{array}[t]{rcl}
\varphi_6 & = & \{2 \cdot \Box\} 
\end{array}$
&
$\begin{array}[t]{rcl}
\ArithTwo(\varphi_6)  & = & 2
\end{array}$\\[0.1cm]
\bottomrule
\end{tabular}
}%end resizebox
%\end{center}
\end{table}

%For $\varphi = (x \vee \neg y \vee z) \wedge (\neg x \vee y \vee z) \wedge(\neg x \vee z) \wedge (\neg y \vee z)$ we have
%$R_x(\varphi) = (\neg y \vee y \vee z \vee z) \wedge (\neg y \vee z \vee z) \wedge (\neg y \vee z)$, and $C_z(C_y(R_x(\varphi)))= (\neg y \vee z) \wedge (\neg y \vee z)$.
\end{exa}

It is well-known that the Davis-Putnam resolution procedure is complete, but we give a proof suitable for our purposes. Let $\varphi[x:=\True]$ denote the result of replacing all occurrences of $x$ in $\varphi$ by $\True$ and all occurrences of $\neg x$ by $\False$. Define $\varphi[x:=\False]$ reversely. Further, let $\exists x \varphi$ be an abbreviation of $\varphi[x:=\True] \vee \varphi[x:=\False]$.
%and $\forall x \varphi$ an abbreviation of $\varphi[x:=\True] \wedge \varphi[x:=\False]$. 
We have:

\begin{lem}\label{lemma:fullressteplemma} Let $x$ be a variable and $\varphi$ a formula in CNF such that $C(x) + C(\neg x) \leq 1$ for every clause $C$. Then $R_x(\varphi) \equiv \exists x \varphi$. \end{lem}
\begin{proof}
Let $C_1,...,C_k$ be the clauses of $\varphi$. We have
\begin{align*}
\exists x \varphi &\equiv \varphi[x:=\True] \vee \varphi[x:=\False]\\
&\equiv \Big(\bigwedge_{i \in [k]} C_i[x:=\True]\Big) \vee \Big(\bigwedge_{j \in [k]} C_j[x:=\False]\Big)\\
&\equiv \bigwedge_{i,j \in [k]}\big(C_i[x:=\True] \vee C_j[x:=\False]\big)\\
&\equiv\bigwedge_{i \in [k], \; x, \neg x \notin C_i} C_i \wedge \bigwedge_{i,j \in [k], \neg x \in C_i, x \in C_j} \big(C_i[x:=\True] \vee C_j[x:=\False]\big)\\
&\equiv R_x(\varphi).
\end{align*}
For the second-to-last equivalence, consider a clause $C_i$ containing neither $x$ nor $\neg x$. Then $C_i \vee C_i$ is a clause of 
$\bigwedge_{i,j \in [k]} \big(C_i[x:=\True] \vee C_j[x:=\False]\big)$, and it subsumes any other clause of the form $C_i \vee C_j$. The first conjunct of the penultimate line contains these clauses. Furthermore, if $C_i$ contains $x$ or if $C_j$ contains $\neg x$, then the disjunction $C_i[x:=\True] \vee C_j[x:=\False]$ is a tautology and can thus be ignored. It remains to consider the pairs $(C_i, C_j)$ of clauses such that $\neg x \in C_i$ and $x \in C_j$. This is the second conjunct. \Qed\end{proof}

\begin{lem}\label{lemma:cleanuplemma} Let $x$ be a variable and $\varphi$ a formula in CNF such that $C(x) + C(\neg x) \leq 2$ for every clause $C$. Then $C_x(\varphi) \equiv \varphi$. \end{lem}
\begin{proof} Since $x \vee \neg x \equiv \True$, a clause containing both $x$ and $\neg x$ is valid and thus can be removed. Furthermore, duplicates of $x$ in a clause can be removed because $x \vee x \equiv x$.\Qed\end{proof}

\begin{thm}\label{theorem:fullresthm} \FullResolution\ is sound and complete. \end{thm}
\begin{proof} Let $\varphi$ be a formula over the variables $x_1,...,x_n$. By Lemmas~\ref{lemma:fullressteplemma} and~\ref{lemma:cleanuplemma}, after termination the algorithm arrives at a formula without variables equivalent to $\exists x_n \cdots \exists x_1 \varphi$. This final formula is equivalent to the truth value of whether $\varphi$ is satisfiable; that is, $\varphi$ is unsatisfiable if{}f the final formula contains the empty clause.\Qed\end{proof}

\section{Constructing Competitive Interactive Protocols for UNSAT}
\label{sec:abstractprotocol}

We consider algorithms for UNSAT that, given a formula, execute a sequence of \emph{macrosteps}. Throughout this section, we use \FullResolution\ as running example.
\begin{defi}
\label{def:macrostep}
A \emph{macrostep} is a partial mapping $M$ that transforms a formula $\varphi$ into a formula $M(\varphi)$ equisatisfiable to $\varphi$.
\end{defi}

The first macrostep is applied to the input formula. The algorithm accepts if the formula returned by the last macrostep is equivalent to $\False$. Clearly, all these algorithms are sound.

\begin{exa}
The macrosteps of \FullResolution\ are $R_x$ and $C_x$ for each variable $x$. On a formula with $n$ variables, \FullResolution\ executes exactly $\frac{n(n+1)}{2}$ macrosteps.
\end{exa}

We present an abstract design framework to obtain competitive interactive protocols for these macrostep-based algorithms.
As in~\cite{LundFKN92,Shamir92,CouillardCEM23}, the framework is based on \emph{arithmetisation} of formulas. Arithmetisations are mappings that assign to a formula a polynomial with integer coefficients. In protocols, Verifier asks Prover to return the result of evaluating polynomials obtained by arithmetising formulas not over the integers, but over a prime field $\F_q$, where $q$ is a sufficiently large prime. An arithmetisation is useful for the design of protocols if the value of the polynomial on a \emph{binary input}, that is, an assignment that assigns $0$ or $1$ to every variable, determines the truth value of the formula under the assignment. We are interested in the following class of arithmetisations, just called  \emph{arithmetisations} for brevity:

\begin{defi}
\label{def:arith}
Let $\Formulas$ and $\Polynomials$ denote the sets of formulas and polynomials over a set of variables. An \emph{arithmetisation} is a mapping $\Arith \colon \Formulas \to \Polynomials$ such that for every formula $\varphi$ and every assignment $\sigma$ to its variables:
\begin{enumerate}[label={(\alph*)}]
\item $\sigma$ satisfies $\varphi$ if{}f $\Pev{\sigma} \Arith(\varphi) = 0$,\footnote{In most papers one requires that $\sigma$ satisfies $\varphi$ if{}f $\Pev{\sigma} \Arith(\varphi) = 1$. Because of our later choice of arithmetisations, we prefer $\Pev{\sigma} \Arith(\varphi) = 0$.} and 
\item $\Pev{\sigma}\Arith(\varphi)\pmod q$ can be computed in time $\O(\Abs{\varphi}\Polylog q)$ for any prime $q$.
\end{enumerate}
\end{defi}

In particular, two formulas $\varphi, \psi$ over the same set of variables are equivalent if and only if for every binary assignment $\sigma$, $\Pev{\sigma} \Arith(\varphi)$ and $\Pev{\sigma} \Arith(\psi)$ are either both zero or both nonzero.

\begin{exa}
\label{ex:arith}
Let $\ArithOne$ be the mapping inductively defined as follows:
$$\begin{array}{l@{\hskip 5mm}l@{\hskip 5mm}l}
\ArithOne(\True) := 0 & \ArithOne(\neg x) := x & \ArithOne(\varphi_1 \wedge \varphi_2) := \ArithOne(\varphi_1) + \ArithOne(\varphi_2) \\[0.2cm]
\ArithOne(\False):= 1 & \ArithOne(x) := 1 - x & \ArithOne(\varphi_1 \vee \varphi_2) := \ArithOne(\varphi_1) \cdot \ArithOne(\varphi_2) .
\end{array}$$
For example, $\ArithOne((x \vee \False) \wedge \neg  x) = ((1-x) \cdot 1) + x = 1$.
It is easy to see that $\ArithOne$ is an arithmetisation in the sense of Definition~\ref{def:arith}. Notice that $\ArithOne$  can map equivalent formulas to different polynomials. For example, $\ArithOne(\neg x)=x$ and $\ArithOne(\neg x \wedge \neg x)=2x$.
\end{exa}

We define when an arithmetisation $\Arith$ is compatible with 
a macrostep $M$.

\begin{defi}\label{def:comp}
Let $\Arith \colon \Formulas \to \Polynomials$ be an arithmetisation and let $M \colon \Formulas \to \Formulas$ be a macrostep. $\Arith$ is \emph{compatible with $M$} if there exists a partial mapping $P_M \colon \Polynomials \to \Polynomials$ and a \emph{pivot variable} $x\in X$ satisfying the following conditions:
\begin{enumerate}[label={(\alph*)}]
\item $P_M$ \emph{simulates} $M$: For every formula $\varphi$ where $M(\varphi)$ is defined, we have $\Arith(M(\varphi)) = P_M(\Arith(\varphi))$.
\item $P_M$ \emph{commutes with partial evaluations}: For every polynomial $p$ and every assignment $\sigma \colon X \setminus \{x\} \to \Z$:
$\Pi_{\sigma}(P_M(p)) = P_M(\Pi_{\sigma}(p))$.
\item $P_M(p\pmod q)=P_M(p)\pmod q$ for any prime $q$.~\footnote{We implicitly extend $P_M$ to polynomials over $\F_q$ in the obvious way: we consider the input $p$ as a polynomial over $\Z$ by selecting the smallest representative in $\N$ for each coefficient, apply $P_M$, and then take the coefficients of the output polynomial modulo $q$.}
\item $P_M$ can be computed in polynomial time.
\end{enumerate}
An arithmetisation $\Arith$ is \emph{compatible} with $\Algorithm$ if it is compatible with every macrostep executed by $\Algorithm$.
\end{defi}

%Figure~\ref{fig:comm} visualises Definition~\ref{def:comp}: 

Graphically, an arithmetisation $\mathcal{A}$ is compatible with $M$ if there exists a mapping $P_M$ such that the following diagram commutes:
%\begin{figure}
\tikzcdset{every label/.append style = {font = \normalsize}}
\[\begin{tikzcd}[sep = large]
	\bullet & \bullet & \bullet & \bullet \\
	\bullet & \bullet & \bullet & \bullet
	\arrow["\Arith", from=1-1, to=1-2]
	\arrow["M", from=1-1, to=2-1]
	\arrow["\Arith", from=2-1, to=2-2]
	\arrow["{P_M}", from=1-2, to=2-2]
	\arrow["{\Pev{\sigma}}", from=1-2, to=1-3]
	\arrow["{\Pev{\sigma}}", from=2-2, to=2-3]
	\arrow["{P_M}", from=1-3, to=2-3]
	\arrow["{\bmod\;q}", from=1-3, to=1-4]
	\arrow["{\bmod\;q}", from=2-3, to=2-4]
	\arrow["{P_M}", from=1-4, to=2-4]
\end{tikzcd}\]
%\caption{Commutative diagram corresponding to Definition~\ref{def:comp}.}
%\label{fig:comm}
%\end{figure}

We can now state and prove the main theorem of this section.

\begin{thm}\label{thm:strategy}
Let $\Algorithm$ be an algorithm for UNSAT and let  $\Arith$ be an arithmetisation compatible with $\Algorithm$ such that for every input $\varphi$
\begin{enumerate}[label={(\alph*)}]
\item\label{wish-a} $\Algorithm$ executes a sequence of $k \in \O(\Poly\Abs{\varphi})$ macrosteps, which compute a sequence 
$\varphi_0, \varphi_1,...,\varphi_k$ of formulas with $\varphi_0=\varphi$,
\item\label{wish-b} $\Arith(\varphi_i)$ has maximum degree at most $d\in\O(\Poly\Abs{\varphi})$, for any $i$, and
\item\label{wish-c} $\Arith(\varphi_k)$ is a constant polynomial such that  $\Abs{\Arith(\varphi_k)}\le 2^{2^{\O(\Abs{\varphi})}}$.
%\item\label{wish-d} $\Arith(\varphi_k)\ne 0$ if{}f $\varphi$ is unsatisfiable, and
%\item\label{wish-c} $\Abs{\Arith(\varphi_k)}\le 2^{2^{\O(\Abs{\varphi})}}$,
\end{enumerate}
Then there is an interactive protocol for UNSAT that is competitive with $\Algorithm$.
\end{thm}

To prove Theorem~\ref{thm:strategy}, we first define a generic interactive protocol for UNSAT depending only on $\Algorithm$ and $\Arith$, and then prove that it satisfies the properties of an interactive proof system: if $\varphi$ is unsatisfiable and Prover is honest, Verifier always accepts; and if $\varphi$ is satisfiable, then Verifier accepts with probability at most $2^{-\Abs{\varphi}}$, regardless of Prover.

\subsection{Interactive Protocol}
The interactive protocol for a given algorithm $\Algorithm$ operates on polynomials over a prime finite field, instead of the integers. Given a prime $q$, we write $\Arith_q(p):=\Arith(p)\pmod q$ for the polynomial over $\F_q$ (the finite field with $q$ elements) that one obtains by taking the coefficients of $\Arith(p)$ modulo $q$.

At the start of the protocol, Prover sends Verifier a prime $q$, and then exchanges messages with Verifier about the values of polynomials over $\F_q$, with the goal of convincing Verifier that $\Arith(\varphi_k)\neq 0$ by showing $\Arith_q(\varphi_k)\neq 0$ instead. The following lemma demonstrates that this is both sound and complete; (a) shows that a dishonest Prover cannot cheat in this way, and (b) shows that an honest Prover can always convince Verifier.

\begin{lem}\label{lemma:primeexists}
Let $\varphi_k$ be the last formula computed by $\Algorithm$.
\begin{enumerate}[label={(\alph*)}]
\item For every prime $q$, we have that $\Arith_q(\varphi_k)\ne 0$ implies that $\varphi$ is unsatisfiable.
\item If $\varphi$ is unsatisfiable, then there exists a prime $q$ s.t.\ $\Arith_q(\varphi_k)\ne 0$.
\end{enumerate}
\end{lem}
\begin{proof}
For every prime $q$, if $\Arith_q(\varphi_k)\ne 0$ then $\Arith(\varphi_k)\ne 0$. For the converse, pick any prime $q$ larger than 
$\Arith(\varphi_k)$. \Qed
%\begin{enumerate}[label={(\alph*)}]
%\item $\Arith_q(\varphi_k)\ne 0$ implies $\Arith(\varphi_k)\ne 0$, which implies that $\varphi$ is unsatisfiable.
%\item Trivial.
%\end{enumerate}\Qed
\end{proof}
Note that we show later that actually Prover can always find a suitable $q$ much smaller than $\Arith(\varphi_k)$ (see Lemma~\ref{lemma:canpickprime}).

We let $\varphi =\varphi_0, \varphi_1, \ldots, \varphi_k$ denote the sequence of formulas computed by $\Algorithm$, and $d$ the bound on the polynomials $\Arith(\varphi_i)$ of Theorem~\ref{thm:strategy}. Observe that the formulas in the sequence can be exponentially larger than $\varphi$, and so Verifier cannot even read them. For this reason, during the protocol Verifier repeatedly sends Prover partial assignments $\sigma$ chosen at random, and Prover sends back to Verifier \emph{claims} about the formulas of the sequence of the form $\Pev{\sigma}\Arith_q(\varphi_i)=w$.  The first claim is about $\varphi_k$, the second about $\varphi_{k-1}$, and so on. Verifier stores the current claim by maintaining variables $i$, $w$, and $\sigma$. The protocol guarantees that the claim about  $\varphi_i$ \emph{reduces} to the  claim about $\varphi_{i-1}$, in the following sense: if a dishonest Prover makes a false claim about $\varphi_i$ but a true claim about $\varphi_{i+1}$, Verifier detects with high probability that the claim about $\varphi_i$ is false and rejects. Therefore, in order to make Verifier accept a satisfiable formula $\varphi$, a dishonest Prover must keep making false claims, and in particular make a false last claim about $\varphi_0=\varphi$.  The protocol also guarantees that a false claim about $\varphi_0$ is always detected by Verifier.

The protocol is described in Table~\ref{tab:prot}.  It presents the steps of Verifier and an \emph{honest} Prover. 

\begin{exa}
In the next section we use the generic protocol of Table~\ref{tab:prot} to give an interactive protocol for $\Algorithm:= \FullResolution$,
using an arithmetisation called $\ArithTwo$. Table~\ref{tab:run} shows a possible run of this protocol on the formula $\varphi$ of Table~\ref{tab:runningexample}. We can already explain the shape of the run, even if $\ArithTwo$ is not defined yet. 

Recall that on input $\varphi$, \FullResolution\ executes six steps, shown on the left column of Table~\ref{tab:runningexample}, that compute the formulas $\varphi_1, \ldots, \varphi_6$. 
Each row of Table~\ref{tab:run} corresponds to a round of the protocol. In round $i$, Prover sends Verifier the polynomial $p_i$ corresponding to the claim $\Pev{\sigma}\Arith_q(\varphi_i)$ (column Honest Prover). Verifier performs a check on the claim (line with $\stackrel{?}{=}$). If the check passes, Verifier updates $\sigma$ and sends it to Prover as the assignment to be used for the next claim.
\end{exa}

\begin{table}[t]
\caption{An interactive protocol for an algorithm for UNSAT describing the behaviour of Verifier and the honest Prover.}
\label{tab:prot}
\medskip
\fbox{%
\begin{minipage}{0.95\textwidth}
\begin{enumerate}
\itemsep1.5mm
\item Prover picks an appropriate prime $q$; i.e.\ a prime s.t. $\Arith_q(\varphi_k)\ne 0$, where $\varphi_k$ is the last formula computed by $\Algorithm$. (The algorithm to compute $q$ is given later.)  
\item Prover sends both $q$ and $\Arith_q(\varphi_k)$ to Verifier. If Prover sends $\Arith_q(\varphi_k)=0$, Verifier rejects. 
\item Verifier sets $i:=k$, $w:=\Arith_q(\varphi_k)$ (sent by Prover in the previous step), and $\sigma$ to an arbitrary assignment. (Since initially $\Arith_q(\varphi_k)$ is a constant, $\sigma$ is irrelevant.) 
\item For each $i=k,...,1$, the claim about $\varphi_i$ is reduced to a claim about $\varphi_{i-1}$:\\[-2.5mm]
\begin{enumerate}[label={4.\arabic*}]
\item Let $x$ denote the pivot variable of $M_i$ and set $\sigma'$ to the partial assignment that is undefined on $x$ and otherwise matches $\sigma$. Prover sends the polynomial $p:=\Pev{\sigma'}\Arith_q(\varphi_{i-1})$, which is a univariate polynomial in $x$.
\item If the degree of $p$ exceeds $d$ or $\pev{x:=\sigma(x)}P_{M_i}(p)\ne w$, Verifier rejects. Otherwise, Verifier chooses an $r\in\F_q$ uniformly at random and updates $w:=\pev{x:=r}p$ and $\sigma(x):=r$.
\end{enumerate}
\item Finally, Verifier checks the claim $\Pev{\sigma}\Arith_q(\varphi_0)=w$ by itself and rejects if it does not hold. Otherwise, Verifier accepts.
\end{enumerate}
\end{minipage}
}
\end{table}

\subsection{The interactive protocol is correct and competitive with $\Algorithm$}

We need to show that the interactive protocol of Table~\ref{tab:prot} is correct and competitive with $\Algorithm$. We do so by means of a sequence of lemmas. Lemmas~\ref{lemma:protocolcorrect}-\ref{lemma:fastprover} bound the error probability of Verifier and the running time of both Prover and Verifier as a function of the prime $q$. Lemma~\ref{lemma:canpickprime} shows that Prover can efficiently compute a suitable prime. The last part of the section combines the lemmas to prove Theorem~\ref{thm:strategy}.

\subsubsection*{Completeness.}
We start by establishing that an honest Prover can always convince Verifier.
\begin{lem}\label{lemma:protocolcomplete}
If $\varphi$ is unsatisfiable and Prover is honest (i.e.\ acts as described in Table~\ref{tab:prot}), then Verifier accepts with probability 1.
\end{lem}
\begin{proof}
We show that Verifier accepts. First we show that  Verifier does not reject in step 2, i.e.\ that $\Arith_q(\varphi_k)\ne0$. Since $\varphi$ is unsatisfiable by assumption, by Definition~\ref{def:macrostep} we have that $\varphi_k$ is unsatisfiable. Then, Definition~\ref{def:arith}(a) implies $\Arith_q(\varphi_k)\ne0$ (note that $\Arith_q(\varphi_k)$ is constant, by Theorem~\ref{thm:strategy}(c)). 

Let us now argue that the claim $w$ Verifier tracks (i.e., the claim given by the current values of the variables) is always true. In step 3, it is initialised with $w:=\Arith_q(\varphi_k)$, so the claim is true at that point. 

Next, let $w$ now be the claim about $\varphi_i$ with $1 \leq i \leq k$, i.e.\ $w= \Pev{\sigma}\Arith_q(\varphi_i)$ holds by induction hypothesis, and Verifier wants to reduce it to a claim about $\varphi_{i-1}$ (performs step 4).
In step 4.2, Verifier checks $\pev{x:=\sigma(x)}P_{M_i}(p)\stackrel{?}{=} w$. As Prover is honest, it sent $p:=\Pev{\sigma'}\Arith_q(\varphi_{i-1})$ in step 4.1; so the check is equivalent to
\begin{align*}
w&\stackrel{?}{=}\pev{x:=\sigma(x)}P_{M_i}(\Pev{\sigma'}\Arith_q(\varphi_{i-1}))&\text{(Definition~\ref{def:comp}(b))}\\
&=\Pev{\sigma}P_{M_i}(\Arith_q(\varphi_{i-1}))&\text{(Definition~\ref{def:comp}(a,c))}\\
&=\Pev{\sigma}\Arith_q(M_i(\varphi_{i-1}))=\Pev{\sigma}\Arith_q(\varphi_i)&
\end{align*}
By induction hypothesis $w= \Pev{\sigma}\Arith_q(\varphi_i)$ holds, and thus Verifier does not reject.

In step 4.2, Verifier also updates the claim about $\varphi_i$ to a claim about $\varphi_{i-1}$. It selects a random number $r$ and sets $w:=\pev{x:=r}p$ and $\sigma(x):=r$. Due to $p=\Pev{\sigma'}\Arith_q(\varphi_{i-1})$, the new claim, $w= \Pev{\sigma}\Arith_q(\varphi_{i-1})$, will hold for all possible values of $r$.

In step 5, we have a claim about $\varphi_0$, i.e.\ $w= \Pev{\sigma}\Arith_q(\varphi_0)$, and the invariant that the claim is true, so Verifier will accept.
\Qed\end{proof}

\subsubsection*{Probability of error.}
Establishing soundness is more involved. The key idea of the proof (which is the same idea as for other interactive protocols) is that for Verifier to accept erroneously, the claim it tracks must at some point be true. However, initially the claim is false. It thus suffices to show that each step of the algorithm is unlikely to turn a false claim into a true one.

\begin{lem}\label{lemma:protocolcorrect}
Let $\Arith,d,k$ as in Theorem~\ref{thm:strategy}. If $\varphi$ is satisfiable, then for any Prover, honest or not, Verifier accepts with probability at most $dk/q \in \O(\Poly(\Abs{\varphi})/q)$.
\end{lem}
\begin{proof}
Let $i\in\{k,...,1\}$, let $\sigma,w$ denote the values of these variables at the beginning of step 4.1 in iteration $i$, and let $\sigma',w'$ denote the corresponding (updated) values at the end of step 4.2.

We say that Prover \emph{tricks} Verifier at iteration $i$ if the claim tracked by Verifier was false at the beginning of step 4 and is true at the end, i.e.\ $\Pev{\sigma}\Arith_q(\varphi_i)\ne w$ and $\Pev{\sigma'}\Arith_q(\varphi_{i-1})=w'$.

The remainder of the proof is split into three parts.
\begin{enumerate}[label={(\alph*)}]
\item If Verifier accepts, it was tricked.
\item For any $i$, Verifier is tricked at iteration $i$ with probability at most $d/q$.
\item Verifier is tricked with probability at most $dk/q\in \O(\Poly(\Abs{\varphi})/q)$.
\end{enumerate}

\smallskip\noindent\textbf{Part (a).}
If $\varphi$ is satisfiable, then so is $\varphi_k$ (Definition~\ref{def:macrostep}), and thus $\Pev{\sigma}\Arith_q(\varphi_k)=0$ (Definition~\ref{def:arith}(a); also note that $\Pev{\sigma}\Arith_q(\varphi_k)$ is constant). Therefore, in step 2 Prover either claims $\Pev{\sigma}\Arith_q(\varphi_k)=0$ and Verifier rejects, or the initial claim in step 3 is false.

If Verifier is never tricked, the claim remains false until step 5 is executed, at which point Verifier will reject. So to accept, Verifier must be tricked.

\smallskip\noindent\textbf{Part (b).}
Let $i\in\{k,...,1\}$ and assume that the claim is false at the beginning of iteration $i$ of step 4. Now there are two cases. If Prover sends the polynomial $p=\Pev{\sigma'}\Arith_q(\varphi_{i-1})$, then, as argued in the proof of Lemma~\ref{lemma:protocolcomplete}, Verifier's check is equivalent to $w\stackrel{?}{=}\Pev{\sigma}\Arith_q(\varphi_i)$, which is the current claim. However, we have assumed that the claim is false, so Verifier would reject. Hence, Prover must send a polynomial $p\ne\Pev{\sigma'}\Arith_q(\varphi_{i-1})$ (of degree at most $d$) to trick Verifier.

By Lemma~\ref{lem:sz}, the probability that Verifier selects an $r$ with $\pev{x:=r}p=\pev{x:=r}\Pev{\sigma'}\Arith_q(\varphi_{i-1})$ is at most $d/q$. Conversely, with probability at least $1-d/q$, the new claim is false as well and Verifier is not tricked in this iteration.

\smallskip\noindent\textbf{Part (c).}
We have shown that the probability that Verifier is tricked in one iteration is at most $d/q$. By union bound, Verifier is thus tricked with probability at most $dk/q$, as there are $k$ iterations. By conditions \ref{wish-a} and~\ref{wish-b} of Theorem~\ref{thm:strategy}, we get $dk/q \in \O(\Poly(\Abs{\varphi})/q)$. \Qed
\end{proof}

\subsubsection*{Running time of Verifier.} 
The next lemma estimates Verifier's running time  in terms of $\Abs{\varphi}$ and $q$. 

\begin{lem}\label{lemma:fastverifier}
Verifier runs in time $\O(\Poly(\Abs{\varphi} \log q))$.
\end{lem}
\begin{proof}
Verifier performs operations on polynomials of degree at most $d$ with coefficients in $\F_q$. So a polynomial can be represented using $d\log q$ bits, and arithmetic operations are polynomial in that representation. Additionally, Verifier needs to execute $P_{M_i}$ for each $i$, which can also be done in polynomial time (Definition~\ref{def:comp}(c)). There are $k\in\O(\Poly\Abs{\varphi})$ iterations.

Finally, Verifier checks the claim $\Pev{\sigma}\Arith_q(\varphi)=w$ for some assignment $\sigma$ and $w\in\F_q$. Definition~\ref{def:arith} ensures that this takes $\O(\Abs{\varphi}\Polylog q)$ time. The overall running time is therefore in $\O(\Poly(\Abs{\varphi}d\log q))$. The final result follows from condition~\ref{wish-b} of Theorem~\ref{thm:strategy}. \Qed
\end{proof}

\subsubsection*{Running time of Prover.} We give a bound on the running time of Prover, excluding the time needed to compute the prime $q$.

\begin{lem}\label{lemma:fastprover}
Assume that $\Arith$ is an arithmetisation satisfying the conditions of Theorem~\ref{thm:strategy}.
Let $T$ denote the time taken by $\Algorithm$ on $\varphi$. The running time of Prover, excluding the time needed to compute the prime $q$, is $\O(T\Poly\Abs{\varphi}\log q))$.
\end{lem}
\begin{proof}
After picking the prime $q$, Prover has to compute $\Pev{\sigma}\Arith_q(\varphi_i)$ for different $i \in [k]$ and assignments $\sigma$. The conditions of Theorem~\ref{thm:strategy} guarantee that this can be done in time $\O(\Abs{\varphi_i}\Polylog q)\subseteq\O(\Abs{\varphi_i}\Poly(\Abs{\varphi}\log q))$. We have $\sum_i\Abs{\varphi_i}\le T$, as $\Algorithm$ needs to write each $\varphi_i$ during its execution. The total running time follows by summing over $i$. \Qed
\end{proof}

\subsubsection*{Computing the prime $q$.} The previous lemmas show the dependence of Verifier's probabitity of error and the running times of Prover and Verifier as a function of $\Abs{\varphi}$ and $q$. Our final lemma gives a procedure for Prover to compute a suitable prime $q$. Together with the previous lemmas, this will easily yield Theorem~\ref{thm:strategy}.

\begin{lem}\label{lemma:canpickprime}
For every $c> 0$ there exists a randomised procedure for Prover to find a prime $q \in 2^{\O(\Abs{\varphi})}$ such that $q \geq 2^{c\Abs{\varphi}}$ and $\Arith_q(\varphi_k) \neq 0$ in expected time $\O(T\Abs{\varphi})$, where $T$ is the running time of $\Algorithm$.
\end{lem}
\begin{proof}
Assume wlog.\ that $c>1$. Prover first runs $\Algorithm$ to compute $\varphi_k$ and then chooses a prime $q$ with $2^{c\Abs{\varphi}}\le q < 2^{c\Abs{\varphi} + 1}$ uniformly at random; thus $q \in 2^{\O(\Abs{\varphi})}$ is guaranteed. If Prover arrives at $\Arith_q(\varphi_k) = 0$, Prover chooses another prime $q$ in the same way, until one is chosen s.t.\ $\Arith_q(\varphi_k)\ne 0$.

Since $\Abs{\Arith(\varphi_k)}\le 2^{2^{\Abs{\varphi}}}$, $\Arith(\varphi_k)$ is divisible by at most $2^{\Abs{\varphi}}$ different primes. Using the prime number theorem, there are $\Omega(2^{c\Abs{\varphi}}/c\Abs{\varphi})$ primes $2^{c\Abs{\varphi}}\le q < 2^{c\Abs{\varphi} + 1}$, so the probability that the picked $q$ divides $\Arith(\varphi_k)$ is $\O(c\Abs{\varphi}/2^{(c-1)\Abs{\varphi}})$.

Therefore, for any $c>1$ this probability is at most, say, $1/2$ for sufficiently large $\Abs{\varphi}$. In expectation, Prover thus needs to test $2$ primes $q$, and each test takes time $\O(\Abs{\varphi_k}\Polylog q)$ (see Definition~\ref{def:arith}(b)), which is in $\O(T\Abs{\varphi})$.
\Qed
\end{proof}

\subsubsection*{Proof of Theorem~\ref{thm:strategy}.} We can now conclude the proof of the theorem.

\medskip\noindent\textit{Completeness} was already proved in Lemma~\ref{lemma:protocolcomplete}.

\medskip\noindent\textit{Soundness.} We need to ensure that the error probability is at most $2^{-\Abs{\varphi}}$. By Lemma~\ref{lemma:protocolcorrect}, the probability $p$ of error satisfies $p\le dk/q$, where $dk\in\O(\Poly(\Abs{\varphi}))$. So there is a $\xi>0$ with $dk\le2^{\xi\Abs{\varphi}}$. Using $c:=1+\xi$ as constant for Lemma~\ref{lemma:canpickprime}, we are done.

\medskip\noindent\textit{Verifier's running time.}  By Lemma~\ref{lemma:fastverifier}, Verifier runs in time $\O(\Poly(\Abs{\varphi} \log q))$.
Using the prime $q \in 2^{\O(\Abs{\varphi})}$ of Lemma~\ref{lemma:canpickprime}, the running time is $\O(\Poly(\Abs{\varphi})$.

\medskip\noindent\textit{Competitivity.} By Lemma~\ref{lemma:fastprover}, Prover runs in time $\O(T \Poly(\Abs{\varphi}\log q))$ plus the time need to compute the prime, which, by Lemma~\ref{lemma:canpickprime}, is in $\O(T\Poly(\Abs{\varphi}))$. Again using $q\in\O(2^{\Abs{\varphi}})$, we find that the protocol is competitive with $\Algorithm$. \Qed

%Let $T$ denote the time taken by $\Algorithm$ on $\varphi$. By Lemma \ref{}We show that Prover runs in $\O(T\Poly\Abs{\varphi}))$. Prover starts by picking a prime $q$ according to Lemma~\ref{lemma:canpickprime} in time $\O(T)$. Afterwards, Prover has to compute $\Pev{\sigma}\Arith_q(\varphi_i)$ for different $i$ and assignments $\sigma$. Definition~\ref{def:arith} guarantees that this can be done in time $\O(\Abs{\varphi_i}\Polylog q)\subseteq\O(\Abs{\varphi_i}\Poly\Abs{\varphi})$. We have $\sum_i\Abs{\varphi_i}\le T$, as $\Algorithm$ needs to write each $\varphi_i$ during its execution. The total running time follows by summing over $i$.

\section{An Interactive Proof System Competitive with the\texorpdfstring{\\}{} Davis-Putnam Resolution Procedure}
\label{sec:competitive}

In order to give an interactive proof system for the Davis-Putnam resolution procedure, it suffices to find an arithmetisation which is compatible with the full $x$-resolution step $R_x$ and the full $x$-cleanup step $C_x$ such that all properties of Theorem~\ref{thm:strategy} are satisfied. In this section, we present such an arithmetisation.
%\begin{itemize}
%\item We find an arithmetisation that is compatible, but not polynomially compatible, with $R_x$ and $C_x$. Applying Theorem \ref{thm:strategy}, we obtain an interactive proof system in which Verifier may run in exponential time. TODO: We don't do this anymore, right?
%\item We tweak this interactive proof system so that Verifier runs in polynomial time. This is done in Section \ref{sec:duplicates}.
%\end{itemize}

\subsection{An arithmetisation compatible with $R_x$ and $C_x$}
\label{subsec:arithtwo}
We find an arithmetisation compatible with both $R_x$ and $C_x$. Let us first see that
the arithmetisation of Example~\ref{ex:arith} does not work.

\begin{exa}
The arithmetisation $\ArithOne$ of Example~\ref{def:arith} is not compatible with $R_x$.
To see this, let $\varphi = (\neg x \vee \neg y) \wedge (x \vee \neg z) \wedge \neg w$. We have $R_x(\varphi) = (\neg y \vee \neg z) \wedge \neg w$, $\ArithOne(R_x(\varphi)) = yz + w$, and $\ArithOne(\varphi) = xy + (1-x)z + w = x(y-z) + z + w$. If $\ArithOne$ were compatible with $R_x$, then there would exist an operation $P_{R_x}$ on polynomials such that $P_{R_x}(x(y-z) + z + w) = yz + w$ by Definition~\ref{def:comp}(a), and from Definition~\ref{def:comp}(b), we get $P_{R_x} (\Pi_\sigma (x(y-z) + z + w)) = \Pi_\sigma (yz + w)$ for all partial assignments $\sigma: \{y,z,w\} \to \mathbb{Z}$. For $\sigma:=\{y\mapsto 1,z\mapsto0,w\mapsto1\}$, it follows that $P_{R_x} (x+1) = 1$, but for $\sigma:=\{y\mapsto2,z\mapsto1,w\mapsto0\}$, it follows that $P_{R_x} (x+1) = 2$, a contradiction.
\end{exa}

We thus present a non-standard arithmetisation.

\begin{defi}\label{def:ArithTwo} The arithmetisation $\ArithTwo$ of a CNF formula $\varphi$ is the recursively defined polynomial
$$\begin{array}{l@{\hskip 5mm}l@{\hskip 5mm}l}
\ArithTwo(\True) := 0 & \ArithTwo(x) := 1 - x & \ArithTwo(\varphi_1 \wedge \varphi_2) := \ArithTwo(\varphi_1) + \ArithTwo(\varphi_2) \\[0.2cm]
\ArithTwo(\False):= 1 & \ArithTwo(\neg x) := x^3 & \ArithTwo(\varphi_1 \vee \varphi_2) := \ArithTwo(\varphi_1) \cdot \ArithTwo(\varphi_2) .
\end{array}$$
\end{defi}

\begin{exa}
The right column of Table~\ref{tab:runningexample} shows the polynomials obtained by applying $\ArithTwo$ to the formulas on the left. 
For example, we have $\ArithTwo(\varphi_5) = \ArithTwo(\neg z \wedge \neg z \wedge z) = 2\ArithTwo(\neg z) +  \ArithTwo(z) =2z^3 + 1 - z$.
\end{exa}

We first prove that $\ArithTwo$ is indeed an arithmetisation.

\begin{prop}\label{proposition:zeroistrue}
For every formula $\varphi$ and every assignment $\sigma: X \to \{0,1\}$ to the variables $X$ of $\varphi$, we have that $\sigma$ satisfies $\varphi$ iff $\Pev{\sigma} \ArithTwo(\varphi) = 0$.
\end{prop}

\begin{proof}
We prove the statement by induction on the structure of $\varphi$.
The statement is trivially true for $\varphi \in \{\True,\False,x,\neg x\}$. For $\varphi = \varphi_1 \vee \varphi_2$, we have
$$ \begin{aligned} &\sigma \text{ satisfies } \varphi \Leftrightarrow \sigma \text{ satisfies } \varphi_1 \vee \varphi_2 \Leftrightarrow \sigma \text{ satisfies } \varphi_1 \text{ or } \sigma \text{ satisfies } \varphi_2\\
\stackrel{IH}{\Leftrightarrow} \: &\Pev{\sigma}\ArithTwo(\varphi_1) = 0 \vee \Pev{\sigma}\ArithTwo(\varphi_2) = 0 \Leftrightarrow \Pev{\sigma}\ArithTwo(\varphi_1) \cdot \Pev{\sigma}\ArithTwo(\varphi_2) = 0\\
\Leftrightarrow \: &\Pev{\sigma}\ArithTwo(\varphi_1 \vee \varphi_2) = 0 \Leftrightarrow \Pev{\sigma}\ArithTwo(\varphi) = 0, \end{aligned} $$
and for $\varphi = \varphi_1 \wedge \varphi_2$, we have
$$ \begin{aligned} &\sigma \text{ satisfies } \varphi \Leftrightarrow \sigma \text{ satisfies } \varphi_1 \wedge \varphi_2 \Leftrightarrow \sigma \text{ satisfies } \varphi_1 \text{ and } \sigma \text{ satisfies } \varphi_2\\
\overset{IH}{\Leftrightarrow} \: &\Pev{\sigma}\ArithTwo(\varphi_1) = 0 \wedge \Pev{\sigma}\ArithTwo(\varphi_2) = 0 \Leftrightarrow \Pev{\sigma}\ArithTwo(\varphi_1) + \Pev{\sigma}\ArithTwo(\varphi_2) = 0\\
\Leftrightarrow \: &\Pev{\sigma}\ArithTwo(\varphi_1 \wedge \varphi_2) = 0 \Leftrightarrow \Pev{\sigma}\ArithTwo(\varphi) = 0. \end{aligned} $$
The equivalence $\Pev{\sigma}\ArithTwo(\varphi_1) = 0 \wedge \Pev{\sigma}\ArithTwo(\varphi_2) = 0 \Leftrightarrow \Pev{\sigma}\ArithTwo(\varphi_1) + \Pev{\sigma}\ArithTwo(\varphi_2) = 0$ holds because $\Pev{\sigma}\ArithTwo(\varphi)$ cannot be negative for binary assignments $\sigma$. \Qed\end{proof}

\subsubsection{$\ArithTwo$ is compatible with $R_x$.}

We exhibit a mapping $\gamma_x \colon \Polynomials \to \Polynomials$ satisfying the conditions of Definition~\ref{def:comp} for the macrostep $R_x$. Recall that $R_x$ is only defined for formulas $\varphi$ in CNF such that $C(x) + C(\neg x) \leq 1$ for every clause $C$. Since arithmetisations of such formulas only have an $x^3$ term, an $x$ term, and a constant term, it suffices to define $\gamma_x$ for polynomials of the form $a_3x^3 + a_1x + a_0$.

\begin{lem}\label{lemma:RxArith} Let $\gamma_x \colon \Polynomials \to \Polynomials$
be the partial mapping defined by $\gamma_x(a_3x^3 + a_1x + a_0) := -a_3a_1 + a_1+a_0$. The mapping $\gamma_x$ witnesses that $\ArithTwo$ is polynomially compatible with the full resolution macrostep $R_x$. \end{lem}
\begin{proof}
We show that $\gamma_x$ satisfies all properties of Definition~\ref{def:comp}. Let $\varphi$ be a formula in CNF such that $C(x) + C(\neg x) \leq 1$ for every clause $C$ (see Definition~\ref{def:fullresstep}). Then $\varphi$ is of the form
$$ \varphi = \Big(\bigwedge_{i \in [k]} x \vee a_i\Big) \wedge \Big(\bigwedge_{j \in [l]} \neg x \vee b_j\Big) \wedge c $$
where $a_i$, $b_j$ are disjunctions not depending on $x$ and $c$ is a conjunction of clauses not depending on $x$. We have
$ R_x(\varphi) = \bigwedge_{i \in [k], \: j \in [l]} (a_i \vee b_j ) \wedge c$.
Now
$$ \begin{aligned} \ArithTwo(\varphi) &= \sum_{i \in [k]} (1-x) a_i + \sum_{j \in [l]} x^3 b_j + c = \sum_{j \in [l]} b_j x^3 - \sum_{i \in [k]} a_i x + \sum_{i \in [k]} a_i + c \end{aligned} $$
and thus
$$ \begin{aligned} \gamma_x(\ArithTwo(\varphi)) &= \Big(\sum_{j \in [l]} b_j\Big) \Big(\sum_{i \in [k]} a_i\Big) - \sum_{i \in [k]} a_i + \sum_{i \in [k]} a_i + c \\
&= \sum_{i \in [k], \: j \in [l]} a_i b_j + c = 
\ArithTwo(R_x(\varphi)). \end{aligned} $$
This proves (a). Since $\gamma_x$ does not depend on variables other than $x$, (b) is also given. (c) and (d) are trivial. \Qed\end{proof}

%We see that a full resolution step of $\varphi$ w.r.t.\ $x$ generates a formula $\psi$ which is not only equivalent to $\exists x \varphi$, but even whose arithmetisation is equal to the arithmetisation of $\exists x \varphi$. Therefore, to arrive at the arithmetisation of $\psi$ from the arithmetisation of $\varphi$, one can simply compute $p_\varphi[x:=0] \cdot p_\varphi[x:=1]$. This is the basis of the verification algorithm.
%
%In the case that not every clause of $\varphi$ depends on $x$, $\varphi$ is of the form $\varphi_1 \wedge \varphi_2$ where $\varphi_1$ only contains clauses depending on $x$ and $\varphi_2$ only contains clauses not depending on $x$. The formula $\psi$ will then have the arithmetisation $p_{\varphi_1}[x:=0] \cdot p_{\varphi_1}[x:=1] + p_{\varphi_2}$.

\subsubsection{$\ArithTwo$ is compatible with $C_x$.} We exhibit a mapping $\delta_x \colon \Polynomials \to \Polynomials$ satisfying the conditions of Definition~\ref{def:comp} for the cleanup macrostep $C_x$. Recall that $C_x$ is only defined for formulas $\varphi$ in CNF such that $C(x) + C(\neg x) \leq 2$ for every clause $C$. Arithmetisations of such formulas are polynomials of degree at most 6 in each variable, and so it suffices to define $\delta_x$ for these polynomials. Additionally, one can see by exhausting all possibilities that $\ArithTwo$ never produces a $x^5$ term due to the restriction $C(x) + C(\neg x) \leq 2$. Hence, $\delta_x$ can ignore the $a_5$ coefficient.

\begin{lem}\label{lemma:CxArith} Let $\delta_x \colon \Polynomials \to \Polynomials$ be the partial mapping defined by 
$$\delta_x(a_6x^6 + a_5x^5 + \cdots  + a_1x + a_0) := (a_6 + a_4 + a_3)x^3 + (a_2 + a_1)x + a_0 . $$
\noindent The mapping $\delta_x$ witnesses that $\ArithTwo$ is polynomially compatible with $C_x$.
\end{lem}
\begin{proof}
We show that $\delta_x$ satisfies all properties of Definition~\ref{def:comp}. We start with (a). Since $\ArithTwo(C \wedge C') = \ArithTwo(C) + \ArithTwo(C')$ for clauses $C, C'$ and $\delta_x(p_1 + p_2) = \delta_x(p_1) + \delta_x(p_2)$, it suffices to show that $\delta_x(\ArithTwo(C)) = \ArithTwo(C_x(C))$ for all clauses $C$ of $\varphi$. Now let $C$ be a clause of $\varphi$. We assume that $C(x) + C(\neg x) \leq 2$ (see Definition~\ref{def:fullresstep}).
\begin{itemize}
    \item If $C(x) + C(\neg x) \leq 1$, then $\delta_x(\ArithTwo(C)) = \ArithTwo(C) = \ArithTwo(C_x(C))$.
    \item If $C = x \vee x \vee C'$, then $\ArithTwo(C) = (1-x)^2 \ArithTwo(C') = (1 - 2x + x^2) \ArithTwo(C')$, so $\delta_x \ArithTwo(C) = (1 - 2x + x) \ArithTwo(C') = (1 - x) \ArithTwo(C') = \ArithTwo(x \vee C') = \ArithTwo(C_x(C))$.
    \item If $C = \neg x \vee \neg x \vee C'$, then $\ArithTwo(C) = x^6 \ArithTwo(C')$, so $\delta_x \ArithTwo(C) = x^3 \ArithTwo(C') = \ArithTwo(\neg x \vee C') = \ArithTwo(C_x(C))$.
    \item If $C = x \vee \neg x \vee C'$, then $\ArithTwo(C) = (1-x)x^3 \ArithTwo(C') = x^3 \ArithTwo(C') - x^4 \ArithTwo(C')$, so $\delta_x \ArithTwo(C) = x^3 \ArithTwo(C') - x^3 \ArithTwo(C') = 0 = \ArithTwo(C_x(C))$.
\end{itemize}
This proves (a). Since $\delta_x$ does not depend on variables other than $x$, (b) is also given. Parts (c) and (d) are trivial.\Qed
\end{proof}

As observed earlier, \FullResolution\ does not remove duplicate clauses; that is, Prover maintains a multiset of clauses that may contain multiple copies of a clause. We show that the number of copies is at most double-exponential in $\Abs{\varphi}$.

\begin{lem}\label{lemma:doubleexp} Let $\varphi$ be the input formula, and let $\varphi_k$ be the last formula computed by \FullResolution. Then $\Arith(\varphi_k) \in 2^{2^{\O(\Abs{\varphi})}}$. \end{lem}
\begin{proof} Let $n_C(\psi)$ be the number of clauses in a formula $\psi$, let $x$ be a variable. Then $n_C(C_x(\psi)) \leq n_C(\psi)$ because a cleanup step can only change or delete clauses. Moreover, $n_C(R_x(\psi)) = n_x n_{\neg x} - n_x - n_{\neg x} + n_C(\psi)$ where $n_x$ and $n_{\neg x}$ are the numbers of clauses in $\psi$ which contain $x$ and $\neg x$, respectively. We get $n_C(R_x(\psi)) \leq (n_x + n_{\neg x})^2 - (n_x + n_{\neg x}) + n_C(\psi)$. Since $n_x + n_{\neg x} \leq n_C(\psi)$, it follows that $n_C(R_x(\psi)) \leq (n_C(\psi))^2$. Now let $n$ be the number of variables. Since $\varphi_k$ is reached after $n$ resolution steps, it follows that $\ArithTwo(\varphi_k) = n_C(\varphi_k) \leq  n_C(\varphi)^{2^n} \in 2^{2^{\O(\Abs{\varphi})}}$. \Qed\end{proof}

\begin{table}[t]
\caption{Run of the instance of the interactive protocol of Table~\ref{tab:prot} for \FullResolution, using the arithmetisation $\ArithTwo$ of Definition~\ref{def:ArithTwo}.}
\label{tab:run}
\begin{center}
\resizebox{\textwidth}{!}{%
\begin{tabular}{|c|l|l|}
\hline
Round & \multicolumn{1}{c|}{Honest Prover}& \multicolumn{1}{c|}{Verifier}
\\ \hline
Initial
& 
\begin{tabular}[t]{l}
$q := 15871$ \\
$p_6 := \ArithTwo_q(\varphi_6) = 2$ \\
send  $q, p_6$
\end{tabular}
&
\begin{tabular}[t]{l}
$w:= p_6 =2$ \\
$\sigma := \{x \mapsto 3, y \mapsto 4, z \mapsto 3\}$\\
send $\sigma$
\end{tabular} 
\\ \hline
$k=6$
&
\begin{tabular}[t]{l}
$\begin{array}[t]{rcl}
\sigma' & := & \{x \mapsto 3, y \mapsto 4\} \\
p_5 & :=& \Pev{\sigma'}(\ArithTwo_q(\varphi_5))  \\
& =  & 2z^3 -z +1
\end{array}$ \\
send $p_5$
\end{tabular} 
&
\begin{tabular}[t]{l}
$\pev{z:=3} \gamma_z(p_5) = \pev{z:=3} 2 \stackrel{?}{=} 2$ \\
$\sigma(z) := 4$ \\
$w := \pev{z:=4} p_5 = 125$ \\
send $\sigma$  \\
\end{tabular}
\\ \hline
$k=5$
&
\begin{tabular}[t]{l}
$\begin{array}[t]{rcl}
\sigma' & := & \{x \mapsto 3, y \mapsto 4\} \\
p_4 & :=& \Pev{\sigma'}(\ArithTwo_q(\varphi_4)) \\
& = & 2z^6 - 3z^4 + 3z^3 + z^2 - 2z + 1
\end{array}$ \\
send $p_4$  
\end{tabular}
&
\begin{tabular}[t]{l}
$\pev{z:=4} \delta_z(p_4) = \pev{z:=4} 2z^3 - z + 1 \stackrel{?}{=}  125$ \\
$\sigma(z) := 2$\\
$w := \pev{z:=2} p_4 = 105$ \\ 
send $\sigma$ \\
\end{tabular}
\\ \hline
$k=4$
&
\begin{tabular}[t]{l}
$\begin{array}[t]{rcl}
\sigma' & := & \{x \mapsto 3, z \mapsto 2\}\\
p_3 & := & \Pev{\sigma'}(\ArithTwo_q(\varphi_3)) \\
& = & 15y^3 - 7y + 7 \\
\end{array}$ \\
send $p_3$
\end{tabular}
&
\begin{tabular}[t]{l}
$\pev{y:=4} \gamma_y(p_3) = \pev{y:=4} 105 \stackrel{?}{=}  105$ \\
$\sigma(y) := 2$\\
$w := \pev{y:=2} p_3 = 113$ \\
send $\sigma$
\end{tabular}
\\ \hline
$k=3$
&
\begin{tabular}[t]{l}
$\begin{array}[t]{rcl}
\sigma' & := & \{x \mapsto 3, y \mapsto 2\}\\
p_2 &:= & \Pev{\sigma'}(\ArithTwo_q(\varphi_2)) \\
& = & 16z^6 - z^3 - 7z + 7 
\end{array}$ \\
send $p_2$  
\end{tabular}
&
\begin{tabular}[t]{l}
$\pev{z:=2} \delta_z(p_2) = \pev{z:=2} 15z^3 - 7z + 7 \stackrel{?}{=}  113$ \\
$\sigma(z) := 3$ \\
$w := \pev{z:=3} p_2 = 11623$ \\
send  $\sigma$ 
\end{tabular}
\\ \hline
$k=2$
&
\begin{tabular}[t]{l}
$\begin{array}[t]{rcl}
\sigma' & := & \{x \mapsto 3, z \mapsto 2\}\\
p_1 &:= & \Pev{\sigma'}(\ArithTwo_q(\varphi_1)) \\
& = & 729y^6 - 27y^4 + 754y^3 - 25y + 25
\end{array}$ \\
send $p_1$
\end{tabular}
&
\begin{tabular}[t]{l}
$\pev{y:=2} \delta_y(p_1) = \pev{y:=2} 1456y^3 - 25y + 25 \stackrel{?}{=} 11623$ \\
$\sigma(y) := 1$ \\
$w := \pev{y:=1} p_1 = 1456$ \\
send $\sigma$
\end{tabular}
\\ \hline
$k=1$
&
\begin{tabular}[t]{l}
$\begin{array}[t]{rcl}
\sigma' & := & \{y \mapsto 1, z \mapsto 2\}\\
p_0 & := & \Pev{\sigma'}(\ArithTwo_q(\varphi_0)) \\
& = &  54x^3 - 27x + 25 \\
\end{array}$ \\
send $p_0$
\end{tabular}
&
\begin{tabular}[t]{l}
$\pev{x:=3} \gamma_x(p_0) = \pev{x:=3} 1456 \stackrel{?}{=} 1456$ \\
$\sigma(x) := 2$ \\
$w := \pev{x:=2} p_0 = 493$ \\
send $\sigma$
\end{tabular}
\\ \hline
Final
& & $\Pev{\sigma} \ArithTwo_q(\varphi) \stackrel{?}{=} 493$\\ \hline
\end{tabular}
} % end resizebox
\end{center}
\end{table}

\begin{prop}\label{prop:fullresolutionprotocol} There exists an interactive protocol for UNSAT that is competitive with \FullResolution.
\end{prop}
\begin{proof}
We show that the $\ArithTwo$ satisfies all properties of Theorem~\ref{thm:strategy}. On an input formula $\varphi$ over $n$ variables, \FullResolution\ executes $n$ resolution steps $R_x$ and $n(n-1)/2$ cleanup steps $C_x$, which gives $n(n+1)/2$ macrosteps in total and proves (a).

Since $\varphi$ does not contain any variable more than once per clause and since cleanup steps w.r.t.\ all remaining variables are applied after every resolution step, resolution steps can only increase the maximum degree of $\ArithTwo(\varphi_i)$ to at most 6 (from 3). Hence the maximum degree of $\ArithTwo(\varphi_i)$ is at most 6 for any $i$, showing (b).

Furthermore, since $R_x(\varphi_i)$ does not contain any occurrence of $x$, and resolution steps are performed w.r.t.\ all variables, $\varphi_k$ does not contain any variables, so $\varphi_k = \{a \cdot \Box\}$ for some $a \in \N$ where $\Box$ is the empty clause. Together with Lemma~\ref{lemma:doubleexp}, (c) follows.
\Qed\end{proof}

Instantiating Theorem~\ref{thm:strategy} with $\ArithTwo$ yields an interactive protocol competitive with \FullResolution. Table~\ref{tab:run} shows a run of this protocol on the formula $\varphi$ of Table~\ref{tab:runningexample}. Initially, Prover runs \FullResolution\ on $\varphi$, computing the formulas $\varphi_1, \ldots, \varphi_6$. Then, during the run of the protocol, it sends to Verifier polynomials of the form $\Pev{\sigma'}\ArithTwo_q(\varphi_{i-1})$ for the assignments $\sigma'$ chosen by Verifier.

\section{Implementation and Evaluation}
\label{sec:impl}
We have implemented  {\FullResolution} as \texttt{icdp} (Interactively Certified Davis-Putnam)~\footnote{\url{https://gitlab.lrz.de/i7/icdp}}, which executes both the Prover and Verifier sides of the interactive protocol. The tool computes a resolution proof of an unsatisfiable SAT instance, and interactively certifies that the proof is correct. The Prover is roughly 1000 lines of C++ code, the Verifier 100 lines, and in total the tool has about 2500 lines of code (all counts excluding blanks and comments). 
We use our implementation to empirically investigate three questions:
\begin{itemize}
\item Interactive vs.\ conventional certification for {\FullResolution}.\\
We compare the resource consumption and communication complexity of (a) interactive certification as carried out by \texttt{icdp} and (b) conventional certification of {\FullResolution}, in which Prover sends Verifier (an external routine) the resolution proof produced by {\FullResolution}, and Verifier sequentially checks the correctness of each resolution step. 
\item \texttt{icdp} vs.\ general-purpose interactive certification. \\
Proofs of the $\textsf{IP}=\textsf{PSPACE}$ theorem provide interactive certification procedures for any \textsf{PSPACE} problem, and so in particular for UNSAT. This raises the question of how this general-purpose procedures compare to \FullResolution. 
\item \texttt{icdp} vs.\ modern DRAT certification. \\
We compare the resource consumption and communication complexity of (a) interactive certification as carried out by \texttt{icdp} and (b) conventional certification as carried out by \texttt{kissat}, a modern SAT solver that emits DRAT certificates of unsatisfiability~\cite{Heule16}. Like resolution proofs, DRAT certificates can be checked by a verifier in linear time and have exponential worst-case size in the size of the formula, but they are more compact in practice\footnote{The related LRAT format~\cite{Cruz-FilipeHHKS17} trades a slight increase in size for faster certification time and would be an interesting comparison as well. We leave this for future work, as DRAT is still the only supported format of almost all modern solvers.}.
\end{itemize}

All experiments are carried out on an AMD Ryzen 9 5900X CPU with 64 GiB of total system memory under Linux, with a timeout of 20 minutes and a memory limit of 48 GiB.

\subsection{Benchmarks and variable order}
\begin{figure}[t]\centering
\ifx\Fullversion\undefined
	\def\w{59.6mm}
	\def\h{2mm}
	\def\hh{4mm}
\else
	\def\w{73mm}
	\def\h{3mm}
	\def\hh{5mm}
\fi
{\footnotesize
        \begin{tabular}{cc}
        \includegraphics[width=\w]{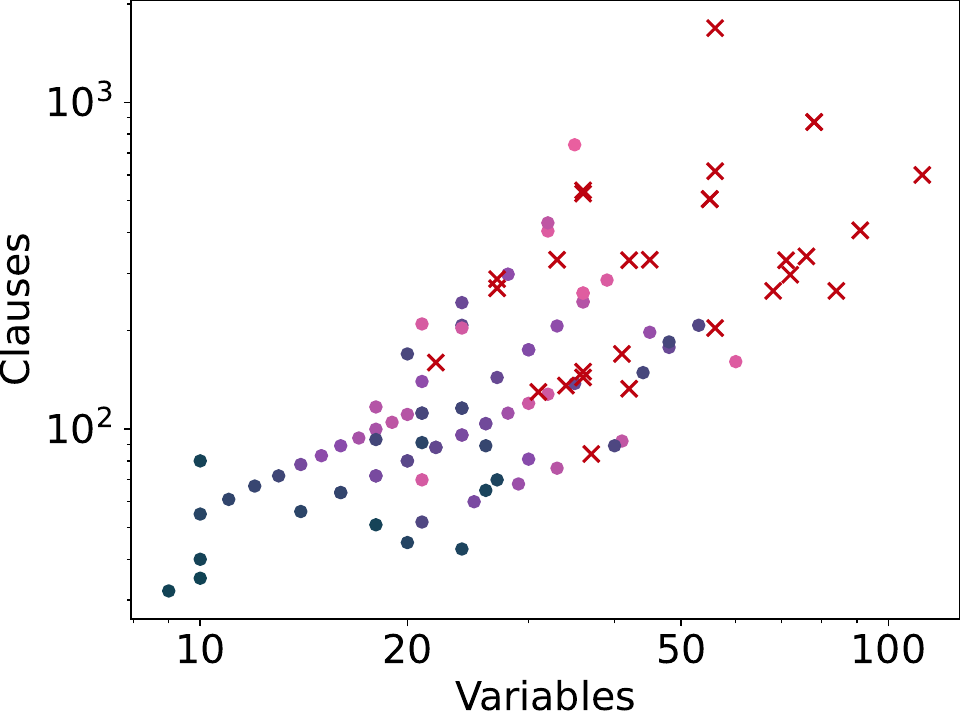}\hspace{\h}
&
        \includegraphics[width=\w]{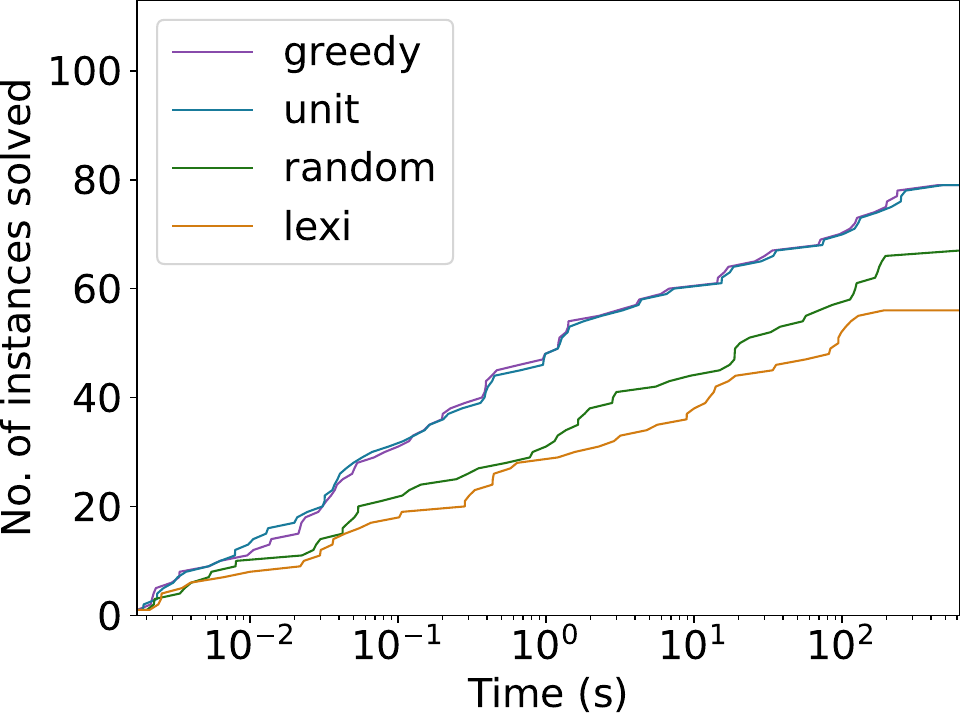} \\
	(a) & (b)
        \end{tabular}
}
        \caption{(a) Size of instances (number of variables and clauses). Instances solved by \texttt{icdp} within 20 minutes using the \texttt{greedy} variable order (see (b)) are shown as dots, the rest as crosses.
                      (b) Number of instances solved by \texttt{icdp} with different variable orderings.}	
\label{fig:newplots}
\end{figure}

To obtain our benchmarks, we first use \texttt{cnfgen}\footnote{\url{https://github.com/MassimoLauria/cnfgen}} to generate 108 unsatisfiable instance formulas from well-known families known to be challenging for many SAT solvers, often encoding combinatorial problems. Table~\ref{tab:families} gives an overview of the families used, and Figure~\ref{fig:newplots}(a) shows the number of variables and clauses of each instance. Dots represent instances solved by \texttt{icdp} within 20 minutes (see below for more details), and crosses represent instances for which \texttt{icdp} times out.

\begin{table}[t]
\caption{Families of formulae used in the benchmark set}
\label{tab:families}
\smallskip
\centering
\begin{tabular}{ll p{30mm} p{70mm}}
Name & Count & Size of largest instance (in KiB) & Description \\ \midrule
bphp               &  5 &  6.1 & Binary pigeon-hole principle \\
cliquecoloring     &  4 &  2.5 & Some graph is both $k$-colourable and has a $(k+1)$-clique \\
count              &  8 & 16.6 & A set of elements can be partitioned into $k$ subsets \\
domset             & 12 &  6.3 & Existence of a dominating set \\
ec                 &  3 &  5.0 & The edges of a graph can be split into two parts s.t.\ each vertex has an equal number of incident edges in both \\
kclique            &  9 &  5.3 & Existence of a clique \\
matching           & 11 &  6.4 & Existence of a perfect matching \\
parity             &  4 &  8.9 & A set can be grouped into pairs \\
php                &  5 &  3.2 & Pigeon-hole principle \\
ram                &  2 &  6.5 & Ramsey numbers \\
randkcnf           & 11 &  1.4 & Random $3$-CNF formulae close to the satisfiability threshold \\
rphp               &  4 &  3.7 & Relativised pigeonhole principle \\
subsetcard         &  9 &  2.3 & 2-coloured edges in a bipartite graph, each colour has a majority in each vertex of the left and right side, respectively \\
tseitin            & 18 &  2.4 & Tseitin formulae \\
vdw                &  3 &  2.3 & Van der Waerden numbers \\
\end{tabular}
\end{table}

The runtime of \texttt{icdp} depends on the variable order. (Recall that Davis-Putnam performs one resolution phase for each variable and its performance depends on the order --- although not its correctness.) We compare four different orders; \texttt{lexi}: the variables are ordered according to the provided instances; \texttt{random}: a variable order is chosen uniformly at random; \texttt{greedy}: at each resolution step, choose the variable that minimises the size of the resulting formula; and \texttt{unit}: if a unit clause exists, pick its variable, and otherwise behave as \texttt{greedy}. The resuts are shown in Figure~\ref{fig:newplots}(b). The \texttt{greedy} order is slightly faster overall, and proves unsatisfiability of 79 of the 108 instances with a timeout of 20 minutes. These are the 79 instances represented as dots in Figure~\ref{fig:newplots}(a), and the ones we use as benchmark set for all subsequent experiments.

\subsection{Interactive vs.\ conventional certification for {\FullResolution}}
We compare the time consumption, memory consumption and communication complexity of interactive certification with \texttt{icdp}, and conventional certification using the resolution proof of {\FullResolution} as certificate. For time and memory consumption, we present results for Prover and Verifier separately. 

\begin{figure}[t]\centering
\ifx\Fullversion\undefined
	\def\w{59.6mm}
	\def\h{2mm}
	\def\hh{4mm}
\else
	\def\w{73mm}
	\def\h{3mm}
	\def\hh{5mm}
\fi
{\footnotesize
        \begin{tabular}{cc}
        \includegraphics[width=\w]{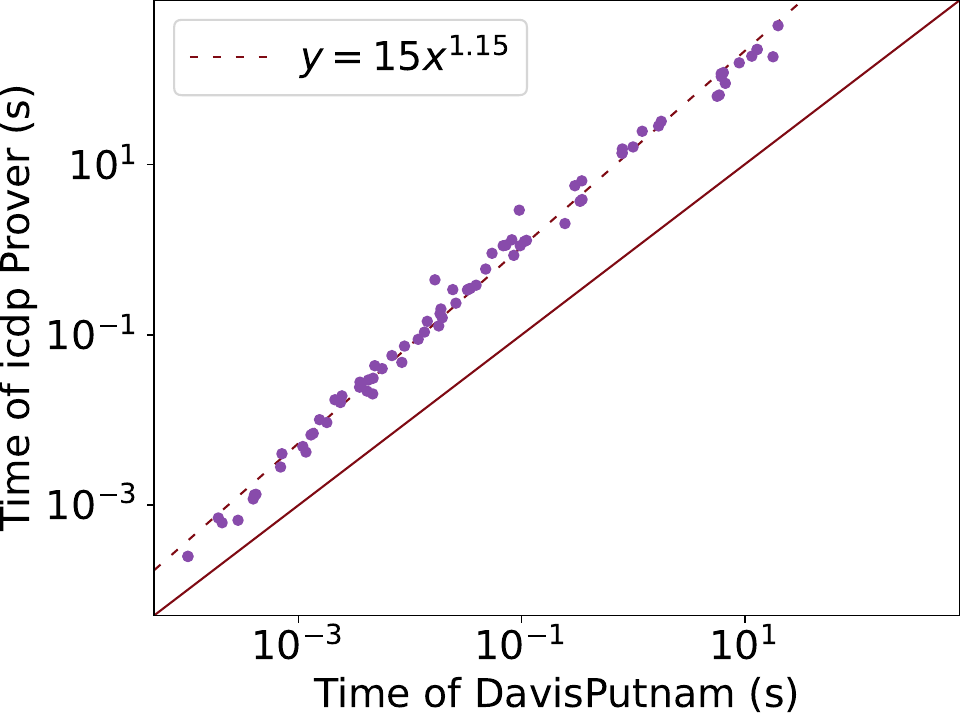}\hspace{\h}
&
        \includegraphics[width=\w]{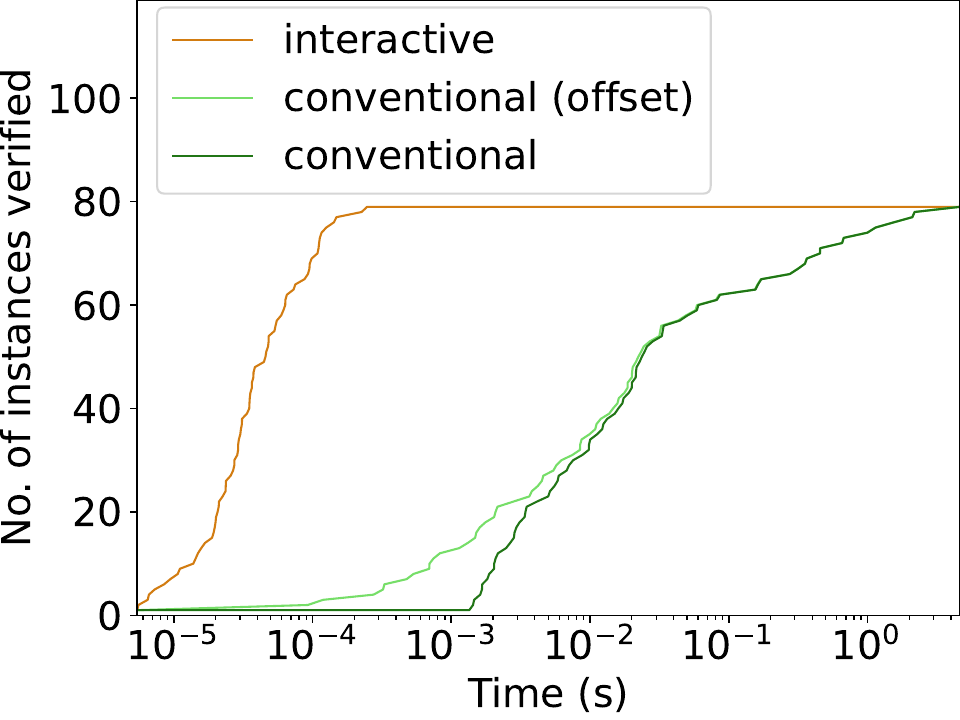} \\
	(a) & (b) 
	
%\\[\hh]
%        \includegraphics[width=\w]{figures/plot1j_survival_checkers.pdf} \hspace{\h}
%&
%        \includegraphics[width=\w]{figures/plot1d_verification_sizes.pdf} \\
%	(c) & (d) \\
        \end{tabular}
}
        \caption{Interactive vs.\ conventional certification: Time consumption of (a) Prover and (b) Verifier, in seconds.
    	}	
\label{fig:time}
        \end{figure}

\smallskip\paragraph{Certification time.}   We first report on the runtime of Prover.  For interactive certification, this is the total time \texttt{icdp}'s Prover needs to \emph{both} solve the instance using {\FullResolution} and interact with Verifier. For conventional certification, Prover's runtime is given by the runtime of {\FullResolution}, because {\FullResolution} already generates a resolution proof that can be used as certificate. 
The results are shown in Figure~\ref{fig:time}(a).  Recall that, by the definition of a competitive protocol, for every formula $\varphi$ the overall time needed by \texttt{icdp}'s Prover is $\O(\Poly(|\varphi|)T(|\varphi|))$, where $T(|\varphi|)$ is the runtime of {\FullResolution} on $\varphi$. Figure~\ref{fig:time}(a) indicates that the polynomial $\Poly(|\varphi|)$ is almost constant.

Now we consider the runtime of Verifier.  For interactive verification, this is the total runtime of \texttt{icdp}'s Verifier. This time is not contiguous, as there are multiple rounds of interaction in the protocol. Intuitively, we measure how much time is spent in \emph{trusted code}, i.e.\ code on which it is sufficient to rely for the correctness of the result. For conventional certification, we let \texttt{icdp} output a resolution proof in the DRAT format used by modern SAT solvers, and define Verifier's runtime as the time taken by the \texttt{DRAT-trim} proof checker to check it~\cite{Heule16}. Due to the design of \texttt{DRAT-trim}, in its default configuration it spends roughly 40ms initialising internal data structures. To reduce this overhead, we adjusted the \texttt{BIGINIT} parameter to 10000 (from 1000000), after which the initialisation takes only 0.5ms. To better compare the scaling behaviour of the tools, we also include times for \texttt{DRAT-trim} that are offset by $T_1-T_2$, where $T_1,T_2$ are the minimum times of \texttt{icdp} and \texttt{DRAT-trim} on any instance, respectively. 

The results are shown in Figure~\ref{fig:time}(b). Since the time used by Verifier for each particular instance is very small, we present a cumulative plot indicating the number of verified instances for a given timeout.  The total certification time for all 79 instances is about four orders of magnitude larger for conventional than for interactive certification.

\begin{figure}[t]\centering
\ifx\Fullversion\undefined
	\def\w{59.6mm}
	\def\h{0.5mm}
	\def\hh{4mm}
\else
	\def\w{73mm}
	\def\h{3mm}
	\def\hh{5mm}
\fi
{\footnotesize
        \includegraphics[width=\w]{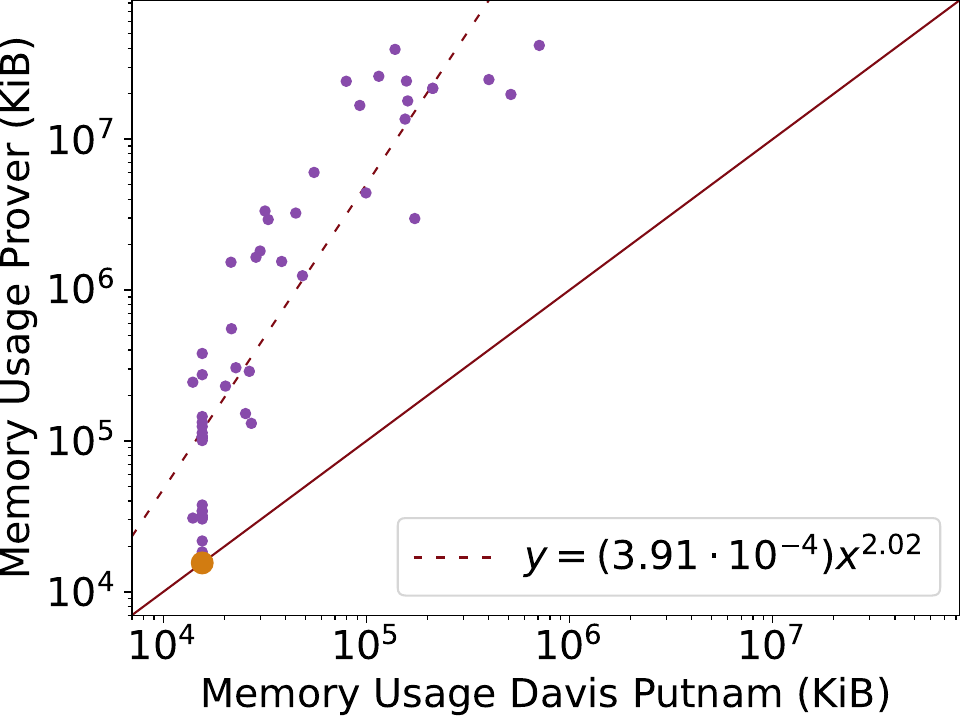}
}
        \caption{Interactive vs.\ conventional certification: Memory usage of Prover, in kilobytes.
    	}	
\label{fig:space}
        \end{figure}

\smallskip\paragraph{Certification space.}  On Prover's side, we plot the memory used by \texttt{icdp}'s Prover to execute the complete protocol against the memory usage of {\FullResolution}. The results are shown in Figure~\ref{fig:space}. The memory required to simply start \texttt{icdp} (without performing any computation) is 15 MiB, and for 27 out of our 79 instances these 15 MiB sufficed for both {\FullResolution} and the complete protocol (orange dot). In the rest of the instances, the memory overhead of \texttt{icdp}'s Prover is quadratic. The reason is that, while \texttt{icdp}'s Prover computes the formulas $\varphi_1, \ldots,  \varphi_k$ (see Table~\ref{tab:prot}) in that order, in the course of certification it needs them in the reverse order, and so \texttt{icdp} stores the formulas $\varphi_1, \ldots,  \varphi_k$. (For convenience they are stored in RAM, but they could also be stored on disk; one could also recompute them, trading time for space.)  On the contrary, at any given time {\FullResolution} only needs to store $\varphi_i$ and $\varphi_{i+1}$ for some $1 \leq i \leq k-1$.

We do not provide quantitative results on the memory usage of Verifier. For both conventional and interactive certification, Verifier needs virtually no memory. 

\begin{figure}[t]\centering
\ifx\Fullversion\undefined
	\def\w{59.6mm}
	\def\h{0.5mm}
	\def\hh{4mm}
\else
	\def\w{73mm}
	\def\h{3mm}
	\def\hh{5mm}
\fi
{\footnotesize
         \includegraphics[width=\w]{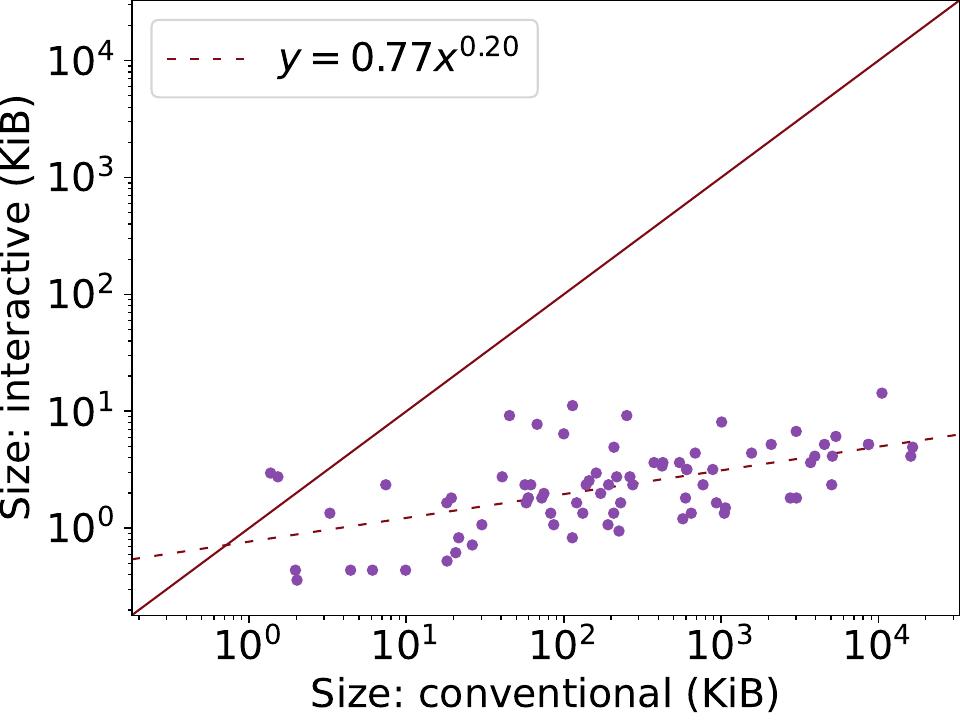} 
}
        \caption{Interactive vs.\ conventional certification: Communication complexity, in kilobytes.
    	}	
\label{fig:commcomplexity}
        \end{figure}

\smallskip\paragraph{Communication complexity.} 
For interactive certification, we count the number of bytes sent by \texttt{icdp}'s Prover to Verifier. For conventional certification, we count the number of bytes of the resolution proof produced by {\FullResolution}. The results are shown in Figure~\ref{fig:commcomplexity}.  We observe that the space used by interactive certification seems to grow like the inverse fifth power of the size of the resolution proof. If this power law extends to larger proofs, then in order to interactively certify a petabyte proof Prover only needs to send Verifier a few kilobytes.

\subsection{\texttt{icdp} vs.\ general-purpose interactive certification}

Proofs of the $\textsf{IP}=\textsf{PSPACE}$ theorem provide interactive certification procedures for any \textsf{PSPACE} problem, and so in particular for UNSAT. This raises the question of how such general-purpose procedures compare in practice with interactive certification using \texttt{icdp}. The answer can be derived from Figure~\ref{fig:newplots}(a). As we mentioned in the introduction, the interactive protocols for UNSAT given by the proofs of $\textsf{IP}=\textsf{PSPACE}$ (or $\textsf{IP} \supseteq \textsf{coNP}$) essentially require that Prover evaluates the formula for all possible truth assignments. Therefore, with a 20-minute timeout for the Prover, and generously assuming that it can check $10^7$ assignments per second, general-purpose Provers can only certify formulas with at most 33 variables, while \texttt{icdp}'s Prover goes beyond 50.

\subsection{\texttt{icdp} vs.\ modern DRAT certification}

Modern SAT solvers emit DRAT certificates of unsatisfiability~\cite{Heule16}. Like resolution proofs, DRAT certificates can be checked by a verifier in linear time and have exponential worst-case size in the size of the formula, but they are more compact in practice\footnote{The related LRAT format~\cite{Cruz-FilipeHHKS17} trades a slight increase in size for faster certification time and would be an interesting comparison as well. We leave this for future work, as DRAT is still the only supported format of almost all modern solvers.}.

We compare the resource consumption of interactive certification with \texttt{icdp} and conventional certification with  \texttt{kissat}\footnote{\url{https://github.com/arminbiere/kissat}}, a state-of-the-art SAT solver. For the latter, we run \texttt{kissat} to generate DRAT certificates, and use \texttt{DRAT-trim} to check their correctness. 

\begin{figure}[t]\centering
\ifx\Fullversion\undefined
	\def\w{59.6mm}
	\def\h{0.5mm}
	\def\hh{4mm}
\else
	\def\w{73mm}
	\def\h{3mm}
	\def\hh{5mm}
\fi
{\footnotesize
         \includegraphics[width=\w]{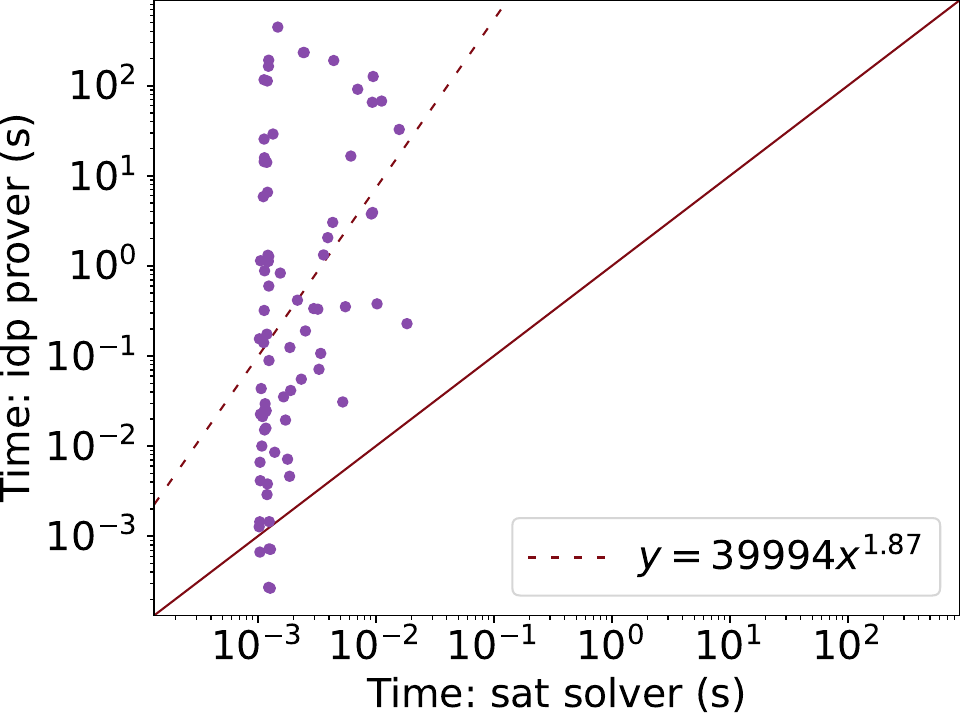} 
}
        \caption{Interactive certification with \texttt{icdp} vs.\ conventional certification with \texttt{kissat} and \texttt{DRAT-trim}: Solving time, in seconds.
    	}	
\label{fig:kissattime}
        \end{figure}

\smallskip\paragraph{Certification time.} On Prover's side, we plot the runtime of \texttt{icdp}'s Prover against  the runtime of \texttt{kissat}.  The results are shown in Figure \ref{fig:kissattime}; \texttt{kissat} 
is about five orders of magnitude faster than \texttt{icdp}.  While a good part of \texttt{kissat}'s advantage may be due to preprocessing and better data structures, we think that the main issue is the low performance of {\FullResolution} itself. As we mentioned in the introduction, {\FullResolution}  was proposed in the 1960s and is not a state-of-the-art procedure.  

\begin{figure}[t]\centering
\ifx\Fullversion\undefined
	\def\w{59.6mm}
	\def\h{0.5mm}
	\def\hh{4mm}
\else
	\def\w{73mm}
	\def\h{3mm}
	\def\hh{5mm}
\fi
{\footnotesize
        \includegraphics[width=\w]{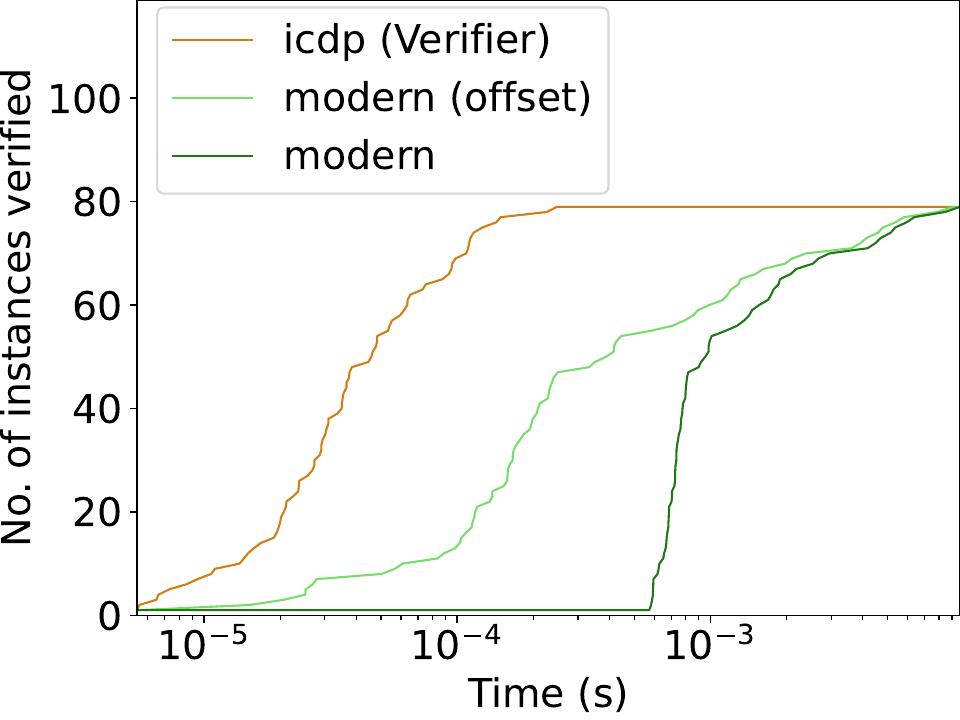} \hspace{\h}
}
        \caption{Interactive certification with \texttt{icdp} vs.\ conventional certification with \texttt{kissat} and \texttt{DRAT-trim}: Verifier's time, in seconds.}	
\label{fig:certtime}
        \end{figure}

Let us now consider the time consumption of Verifier. A caveat is that, since \texttt{icdp} can only certify formulas it has solved, we can only perform a comparison on formulas that are very small for the standards of modern SAT solvers.  We compare the time that \texttt{DRAT-trim}---a standard tool for checking DRAT certificates---needs to check the DRAT certificates produced by \texttt{kissat} with the runtime of \texttt{icdp}'s Verifier. Time is measured internally, using the same high-resolution clock. Figure~\ref{fig:certtime} shows a survival plot of the number of instances checked by Verifier within a given time. As described above, for \texttt{DRAT-trim} we also provide times that are offset. Overall, \texttt{icdp}'s Verifier checks roughly one order of magnitude faster than \texttt{DRAT-trim}, even when the times are offset.  For example, the longest time of \texttt{DRAT-trim} is 347ms, while \texttt{icdp}'s Verifier takes at most 0.2ms on the benchmarked instances\footnote{Recall that certification times may be faster for the LRAT format, and so the advantage of interactive certification with respect to time might disappear in a comparison with LRAT certification.}. This result is likely due to the theoretical exponential speedup of Verifier compared to conventional certification.

\smallskip\paragraph{Certification space.} We omit the experimental results, because they do not add insight: On Prover's side, we have the same situation as for time consumption: \texttt{icdp}'s Prover consumes much more memory than \texttt{kissat}. On Verifier's side, the memory consumption of both \texttt{icdp}'s Verifier and \texttt{DRAT-trim} are negligible.

\begin{figure}[t]\centering
\ifx\Fullversion\undefined
	\def\w{59.6mm}
	\def\h{0.5mm}
	\def\hh{4mm}
\else
	\def\w{73mm}
	\def\h{3mm}
	\def\hh{5mm}
\fi
{\footnotesize
        \includegraphics[width=\w]{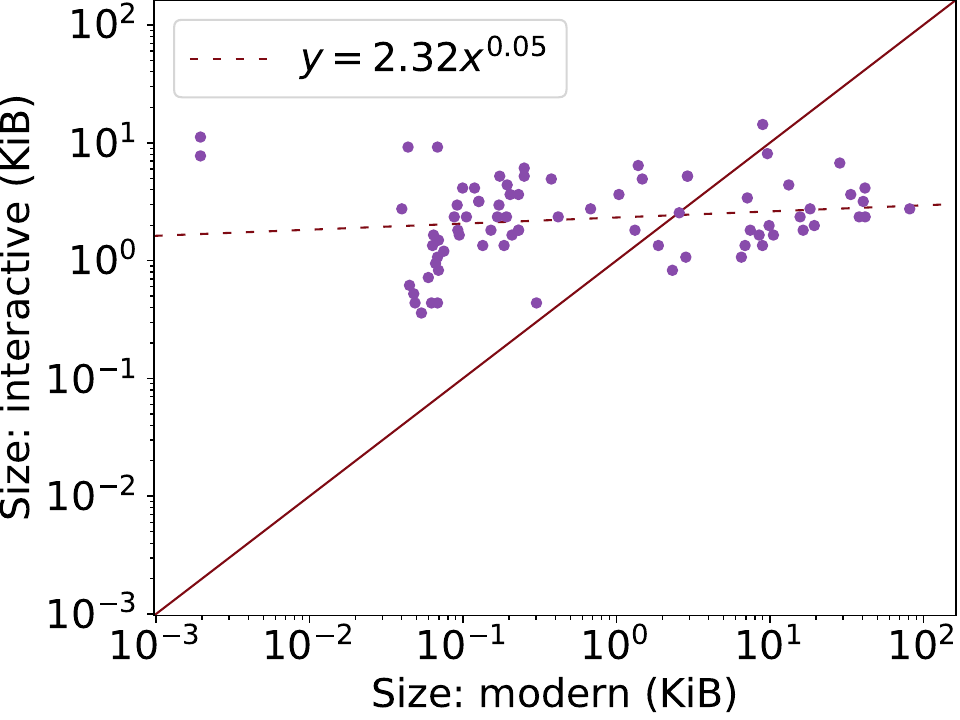} \\
}
       \caption{Interactive certification with \texttt{icdp} vs.\ conventional certification with \texttt{kissat} and \texttt{DRAT-trim}:  Communication complexity, in kilobytes.}	
\label{fig:commcomplexitykissat}
        \end{figure}

\smallskip\paragraph{Communication complexity.} We compare the length in bytes of the DRAT certificate produced by \texttt{kissat} and the number of bytes sent by \texttt{icdp}'s Prover to Verifier during the execution of \texttt{icdp}. The results are shown in Figure~\ref{fig:commcomplexitykissat}. While DRAT certificates are smaller for many instances,  interactive certification scales better, and outperforms conventional certification in the larger instances. In particular, the size of the interactive certificates grows even more slowly in the size of the DRAT certificate than in the size of the resolution proof\footnote{DRAT certificates are more compact than LRAT ones, and so the advantage of interactive certification with respect to certificate size should also manifest itself in an experiment with LRAT certificates.}. Again, this result is due to the theoretical complexity differences between interactive proofs with guaranteed polynomial time for Verifier and conventional certification with exponential-size certificates in the worst case.

\subsection{Final summary}
If we consider {\FullResolution} in a distributed setting, where Prover and Verifier run in different machines, interactive certification exhibits clear advantages w.r.t.\ Verifier's time consumption and communication complexity: both improve by several orders of magnitude with respect to conventional certification. The price to pay on Prover's side is an almost linear time overhead and a quadratic memory overhead.

Our experiments on interactive certification for {\FullResolution} vs.\ conventional certification in \texttt{kissat} indicate that, while the advantages in Verifier's time consumption and communication complexity still remain, the price to pay in Prover's runtime is very high, due to the much higher solving time.

\section{Conclusions}
\label{sec:conclusions}

We have presented the first interactive proof system for the Davis-Putnam resolution procedure, one of the oldest algorithms for UNSAT producing a certificate. We have derived it in a systematic way using a procedure to obtain interactive proof systems from arithmetisations satisfying a few commutativity properties. We have shown that, while standard arithmetisations do not satisfy these properties, non-standard arithmetisations do.

Our work shows that interactive certification with small Prover overhead exist not only for brute-force algorithms, but for more sophisticated ones. At the same time,  Davis-Putnam is several orders of magnitude slower than current techniques, like CDCL. Whether the results we have obtained for Davis-Putnam can be extended to these techniques remains open.

Lovasz \etal\ have shown that given a refutation by the Davis-Putnam resolution procedure, one can extract a multi-valued decision diagram, polynomial in the size of the refutation, in which the path for a given truth assignment leads to a clause false under that assignment (that is, to a clause witnessing that the assignment does not satisfy the formula)~\cite{LovaszNNW95}. This suggests a possible connection between our work and the work of Couillard \etal\ in~\cite{CouillardCEM23}. As mentioned in the introduction,~\cite{CouillardCEM23} presents an interactive proof system competitive with the algorithm for UNSAT that iteratively constructs a BDD for the formula (starting at the leaves of its syntax tree, and moving up at each step), and returns ``unsatisfiable'' if{}f the BDD for the root of the tree only contains the node $0$. We conjecture that a future version of our systematic derivation technique could subsume both~\cite{CouillardCEM23} and this paper.

\paragraph{Acknowledgments.} We thank the anonymous reviewers for their comments and Albert Atserias for helpful discussions.

\vspace*{8pt} % Add space so license information gets printed at the very end
\bibliographystyle{alphaurl}
\bibliography{lmcs-long}

  %% the following bibliography is gererated manually for the sake of brevity
  %% only; please use a separate .bib file in your submission

\end{document}

%% file: packages.tex
\usepackage{tikz}
\usetikzlibrary{automata, positioning, arrows, shapes, arrows.meta}
\usepackage{quiver}

\usepackage{thm-restate}
\usepackage{graphicx}
\usepackage{xcolor}
\usepackage{xspace}
\usepackage{amssymb}
\usepackage[most]{tcolorbox}
\usepackage{wrapfig}
\usepackage{multirow}
\usepackage[strings]{underscore}
\usepackage{booktabs}
\usepackage{multicol}
\usepackage{pbox}
\usepackage{enumitem}

\usepackage{algorithm}
\usepackage[noend]{algpseudocode}
\algrenewcommand\algorithmicindent{1.2em}
\algnewcommand\Input{\item[\textbf{Input:}]}
\algnewcommand\Output{\item[\textbf{Output:}]}
\algnewcommand\Effect{\item[\textbf{Effect:}]}
\allowdisplaybreaks

%% file: chairmacros.tex
\usepackage{xcolor}
\usepackage{stmaryrd}

\newcommand{\N}{\mathbb{N}}

\newcommand{\Z}{\mathbb{Z}}

\def\O{\mathcal{O}}

\newcommand{\NP}{\mathsf{NP}}
\newcommand{\coNP}{\mathsf{coNP}}
\newcommand{\PSPACE}{\textnormal{\textsf{PSPACE}}}
\newcommand{\IP}{\textnormal{\textsf{IP}}}

\newcommand{\Abs}[1]{\mathopen|#1\mathclose|}

\definecolor{nicebg}{HTML}{f6f0e4}
\definecolor{nicered}{HTML}{7f0a13}
\definecolor{nicebgred}{HTML}{f2e7e8}
\definecolor{niceblue}{HTML}{104354}
\definecolor{nicebgblue}{HTML}{e8edee}
\definecolor{nicegreen}{HTML}{217516}
\definecolor{nicebggreen}{HTML}{e9f1e8}
\definecolor{nicepurple}{HTML}{884bab}
\definecolor{nicebgpurple}{HTML}{f3edf7}
\definecolor{niceorange}{HTML}{d27c11}
\definecolor{nicebgorange}{HTML}{fbf2e8}
\definecolor{nicepink}{HTML}{e95f9f}
\definecolor{nicebgpink}{HTML}{fdeff6}
\definecolor{niceredlight}{HTML}{c9888d}
\definecolor{nicebluelight}{HTML}{78a4b8}
\definecolor{nicegreenlight}{HTML}{76de68}
\definecolor{nicepurplelight}{HTML}{bc87db}
\definecolor{niceredbright}{HTML}{bd0310}
\definecolor{nicebgredbright}{HTML}{f9e6e8}
\definecolor{nicebluebright}{HTML}{197b9b}
\definecolor{nicebgbluebright}{HTML}{e8f2f5}

%% file: notation.tex
\newcommand{\F}{\mathbb{F}}
\renewcommand{\O}{\mathcal{O}}
\newcommand{\Poly}{\operatorname{poly}}
\newcommand{\Polylog}{\operatorname{polylog}}

\newcommand{\pev}[1]{\pi_{[{#1}]}}
\newcommand{\Pev}[1]{{\mathrm{\Pi}}_{{#1}}}

\newcommand{\Gets}{:=}
\newcommand{\Alg}[1]{\textnormal{\textsc{#1}}}

\newcommand{\BDDSolver}{\Alg{BDDSolver}}

\newcommand{\FullResolution}{\Alg{DavisPutnam}}
\newcommand{\res}[1]{\textit{Res}_{#1}}
\newcommand{\Arith}{\mathcal{A}}
\newcommand{\Formulas}{\mathcal{F}}
\newcommand{\Polynomials}{\mathcal{P}}

\newcommand{\ArithOne}{\mathcal{A}}
\newcommand{\ArithTwo}{\mathcal{B}}

\newcommand{\True}{\textit{true}}
\newcommand{\False}{\textit{false}}

\newcommand{\Algorithm}{\textit{Alg}}

\renewcommand{\ldots}{...}

\tikzstyle{cnode}=[ellipse,draw,thick,minimum width=2em,minimum height=3ex,fill=nicebgblue]
\tikzstyle{dnode}=[circle,draw,thick,minimum size= 0pt,inner sep=1pt, outer sep=0pt, fill=nicebgredbright]
\tikzstyle{pnode}=[ellipse,draw,thick,fill=nicebgblue]
\tikzstyle{fml}=[rectangle,draw=nicegreen!50!white,fill=nicebggreen,rounded corners,inner sep=1.5mm]
\tikzstyle{poly}=[inner sep=0mm]

\newsavebox{\ProblemBox}
\newlength{\ProblemLength}
\newcommand{\Problem}[4][]{
  \begingroup
  \def\Arg{#1}
  \ifx\Arg\empty\def\Arg{#2}\fi
  \begin{center}
    \savebox{\ProblemBox}{\textsf{#2}}%
    \settowidth{\ProblemLength}{\usebox{\ProblemBox}}%
    \usebox{\ProblemBox}\hspace{1mm}
    \begin{tabular}{ll}
      \textbf{Input}& #3\\
      \textbf{Output}& #4\\
    \end{tabular}
  \end{center}
  \endgroup
}

\newcommand{\Out}{\operatorname{\mathsf{out}}}
\newcommand{\Vtime}{\operatorname{\mathsf{vtime}}}
\newcommand{\Ptime}{\operatorname{\mathsf{ptime}}}

\newcommand{\phii}[1]{\varphi_{#1}}
\newcommand{\etal}{et al.}

%% file: lmcs-long.bib
@incollection{BussN21,
  author       = {Sam Buss and
                  Jakob Nordstr{\"{o}}m},
  editor       = {Armin Biere and
                  Marijn Heule and
                  Hans van Maaren and
                  Toby Walsh},
  title        = {Proof Complexity and {SAT} Solving},
  booktitle    = {Handbook of Satisfiability - Second Edition},
  series       = {Frontiers in Artificial Intelligence and Applications},
  volume       = {336},
  pages        = {233--350},
  publisher    = {{IOS} Press},
  year         = {2021},
  doi          = {10.3233/FAIA200990},
}

@article{BussT88,
  author       = {Samuel R. Buss and
                  Gy{\"{o}}rgy Tur{\'{a}}n},
  title        = {Resolution Proofs of Generalized Pigeonhole Principles},
  journal      = {Theor. Comput. Sci.},
  volume       = {62},
  number       = {3},
  pages        = {311--317},
  year         = {1988},
  doi          = {10.1016/0304-3975(88)90072-2},
}

@article{Haken85,
  author       = {Armin Haken},
  title        = {The Intractability of Resolution},
  journal      = {Theor. Comput. Sci.},
  volume       = {39},
  pages        = {297--308},
  year         = {1985},
  noUrlBecauseOfDoi          = {https://doi.org/10.1016/0304-3975(85)90144-6},
  doi          = {10.1016/0304-3975(85)90144-6},
  bibsource    = {dblp computer science bibliography, https://dblp.org}
}

@article{LovaszNNW95,
  author       = {L{\'{a}}szl{\'{o}} Lov{\'{a}}sz and
                  Moni Naor and
                  Ilan Newman and
                  Avi Wigderson},
  title        = {Search Problems in the Decision Tree Model},
  journal      = {{SIAM} J. Discret. Math.},
  volume       = {8},
  number       = {1},
  pages        = {119--132},
  year         = {1995},
  noUrlBecauseOfDoi          = {https://doi.org/10.1137/S0895480192233867},
  doi          = {10.1137/S0895480192233867},
  bibsource    = {dblp computer science bibliography, https://dblp.org}
}

@inproceedings{HeuleHKW17,
  author       = {Marijn Heule and
                  Warren A. Hunt Jr. and
                  Matt Kaufmann and
                  Nathan Wetzler},
  editor       = {Mauricio Ayala{-}Rinc{\'{o}}n and
                  C{\'{e}}sar A. Mu{\~{n}}oz},
  title        = {Efficient, Verified Checking of Propositional Proofs},
  booktitle    = {Interactive Theorem Proving - 8th International Conference, {ITP}
                  2017, Bras{\'{\i}}lia, Brazil, September 26-29, 2017, Proceedings},
  series       = {Lecture Notes in Computer Science},
  volume       = {10499},
  pages        = {269--284},
  publisher    = {Springer},
  year         = {2017},
  noUrlBecauseOfDoi          = {https://doi.org/10.1007/978-3-319-66107-0\_18},
  doi          = {10.1007/978-3-319-66107-0\_18},
  bibsource    = {dblp computer science bibliography, https://dblp.org}
}

@article{HeuleKM16,
  author       = {Marijn J. H. Heule and
                  Oliver Kullmann and
                  Victor W. Marek},
  title        = {Solving and Verifying the boolean Pythagorean Triples problem via
                  Cube-and-Conquer},
  journal      = {CoRR},
  volume       = {abs/1605.00723},
  year         = {2016}
}

@article{Shamir92,
  author    = {Adi Shamir},
  title     = {{IP} = {PSPACE}},
  journal   = {J. {ACM}},
  volume    = {39},
  number    = {4},
  pages     = {869--877},
  year      = {1992},
  noUrlBecauseOfDoi       = {https://doi.org/10.1145/146585.146609},
  doi       = {10.1145/146585.146609},
  bibsource = {dblp computer science bibliography, https://dblp.org}
}

@inproceedings{BarbosaRKLNNOPV22,
  author    = {Haniel Barbosa and
               Andrew Reynolds and
               Gereon Kremer and
               Hanna Lachnitt and
               Aina Niemetz and
               Andres N{\"{o}}tzli and
               Alex Ozdemir and
               Mathias Preiner and
               Arjun Viswanathan and
               Scott Viteri and
               Yoni Zohar and
               Cesare Tinelli and
               Clark W. Barrett},
  editor    = {Jasmin Blanchette and
               Laura Kov{\'{a}}cs and
               Dirk Pattinson},
  title     = {Flexible Proof Production in an Industrial-Strength {SMT} Solver},
  booktitle = {Automated Reasoning - 11th International Joint Conference, {IJCAR}
               2022, Haifa, Israel, August 8-10, 2022, Proceedings},
  series    = {Lecture Notes in Computer Science},
  volume    = {13385},
  pages     = {15--35},
  publisher = {Springer},
  year      = {2022},
  noUrlBecauseOfDoi       = {https://doi.org/10.1007/978-3-031-10769-6\_3},
  doi       = {10.1007/978-3-031-10769-6\_3},
  bibsource = {dblp computer science bibliography, https://dblp.org}
}

@inproceedings{Babai,
  author    = {L{\'{a}}szl{\'{o}} Babai},
  editor    = {Robert Sedgewick},
  title     = {Trading Group Theory for Randomness},
  booktitle = {Proceedings of the 17th Annual {ACM} Symposium on Theory of Computing,
               May 6-8, 1985, Providence, Rhode Island, {USA}},
  pages     = {421--429},
  publisher = {{ACM}},
  year      = {1985},
  noUrlBecauseOfDoi       = {https://doi.org/10.1145/22145.22192},
  doi       = {10.1145/22145.22192},
  bibsource = {dblp computer science bibliography, https://dblp.org}
}

@inproceedings{GoldwasserMicaliRackoff,
  author    = {Shafi Goldwasser and
               Silvio Micali and
               Charles Rackoff},
  editor    = {Robert Sedgewick},
  title     = {The Knowledge Complexity of Interactive Proof-Systems (Extended Abstract)},
  booktitle = {Proceedings of the 17th Annual {ACM} Symposium on Theory of Computing,
               May 6-8, 1985, Providence, Rhode Island, {USA}},
  pages     = {291--304},
  publisher = {{ACM}},
  year      = {1985},
  noUrlBecauseOfDoi       = {https://doi.org/10.1145/22145.22178},
  doi       = {10.1145/22145.22178},
  bibsource = {dblp computer science bibliography, https://dblp.org}
}

@article{LundFKN92,
  author    = {Carsten Lund and
               Lance Fortnow and
               Howard J. Karloff and
               Noam Nisan},
  title     = {Algebraic Methods for Interactive Proof Systems},
  journal   = {J. {ACM}},
  volume    = {39},
  number    = {4},
  pages     = {859--868},
  year      = {1992},
  noUrlBecauseOfDoi       = {https://doi.org/10.1145/146585.146605},
  doi       = {10.1145/146585.146605},
  bibsource = {dblp computer science bibliography, https://dblp.org}
}

@inproceedings{HenzingerJMNSW02,
  author       = {Thomas A. Henzinger and
                  Ranjit Jhala and
                  Rupak Majumdar and
                  George C. Necula and
                  Gr{\'{e}}goire Sutre and
                  Westley Weimer},
  editor       = {Ed Brinksma and
                  Kim Guldstrand Larsen},
  title        = {Temporal-Safety Proofs for Systems Code},
  booktitle    = {Computer Aided Verification, 14th International Conference, {CAV}
                  2002,Copenhagen, Denmark, July 27-31, 2002, Proceedings},
  series       = {Lecture Notes in Computer Science},
  volume       = {2404},
  pages        = {526--538},
  publisher    = {Springer},
  year         = {2002},
  noUrlBecauseOfDoi          = {https://doi.org/10.1007/3-540-45657-0\_45},
  doi          = {10.1007/3-540-45657-0\_45},
  bibsource    = {dblp computer science bibliography, https://dblp.org}
}

@inproceedings{Namjoshi01,
  author       = {Kedar S. Namjoshi},
  editor       = {G{\'{e}}rard Berry and
                  Hubert Comon and
                  Alain Finkel},
  title        = {Certifying Model Checkers},
  booktitle    = {Computer Aided Verification, 13th International Conference, {CAV}
                  2001, Paris, France, July 18-22, 2001, Proceedings},
  series       = {Lecture Notes in Computer Science},
  volume       = {2102},
  pages        = {2--13},
  publisher    = {Springer},
  year         = {2001},
  noUrlBecauseOfDoi          = {https://doi.org/10.1007/3-540-44585-4\_2},
  doi          = {10.1007/3-540-44585-4\_2},
  bibsource    = {dblp computer science bibliography, https://dblp.org}
}

@inproceedings{Necula97,
  author       = {George C. Necula},
  editor       = {Peter Lee and
                  Fritz Henglein and
                  Neil D. Jones},
  title        = {Proof-Carrying Code},
  booktitle    = {Conference Record of POPL'97: The 24th {ACM} {SIGPLAN-SIGACT} Symposium
                  on Principles of Programming Languages, Papers Presented at the Symposium,
                  Paris, France, 15-17 January 1997},
  pages        = {106--119},
  publisher    = {{ACM} Press},
  year         = {1997},
  noUrlBecauseOfDoi          = {https://doi.org/10.1145/263699.263712},
  doi          = {10.1145/263699.263712},
  bibsource    = {dblp computer science bibliography, https://dblp.org}
}

@incollection{Heule21,
  author    = {Marijn J. H. Heule},
  editor    = {Armin Biere and
               Marijn Heule and
               Hans van Maaren and
               Toby Walsh},
  title     = {Proofs of Unsatisfiability},
  booktitle = {Handbook of Satisfiability - Second Edition},
  series    = {Frontiers in Artificial Intelligence and Applications},
  volume    = {336},
  pages     = {635--668},
  publisher = {{IOS} Press},
  year      = {2021},
  noUrlBecauseOfDoi       = {https://doi.org/10.3233/FAIA200998},
  doi       = {10.3233/FAIA200998},
  bibsource = {dblp computer science bibliography, https://dblp.org}
}

@inproceedings{CouillardCEM23,
  author       = {Eszter Couillard and
                  Philipp Czerner and
                  Javier Esparza and
                  Rupak Majumdar},
  editor       = {Constantin Enea and
                  Akash Lal},
  title        = {Making $\textsf{IP}=\textsf{PSPACE}$ Practical: Efficient Interactive Protocols
                  for {BDD} Algorithms},
  booktitle    = {Computer Aided Verification - 35th International Conference, {CAV}
                  2023, Paris, France, July 17-22, 2023, Proceedings, Part {III}},
  series       = {Lecture Notes in Computer Science},
  volume       = {13966},
  pages        = {437--458},
  publisher    = {Springer},
  year         = {2023},
  noUrlBecauseOfDoi          = {https://doi.org/10.1007/978-3-031-37709-9\_21},
  doi          = {10.1007/978-3-031-37709-9\_21},
  bibsource    = {dblp computer science bibliography, https://dblp.org}
}

@article{DavisP60,
  author       = {Martin Davis and
                  Hilary Putnam},
  title        = {A Computing Procedure for Quantification Theory},
  journal      = {J. {ACM}},
  volume       = {7},
  number       = {3},
  pages        = {201--215},
  year         = {1960},
  noUrlBecauseOfDoi          = {https://doi.org/10.1145/321033.321034},
  doi          = {10.1145/321033.321034},
  bibsource    = {dblp computer science bibliography, https://dblp.org}
}

@book{Harrison09,
  author       = {John Harrison},
  title        = {Handbook of Practical Logic and Automated Reasoning},
  publisher    = {Cambridge University Press},
  year         = {2009},
  isbn         = {978-0-521-89957-4},
  bibsource    = {dblp computer science bibliography, https://dblp.org}
}

@article{Heule16,
  author       = {Marijn J. H. Heule},
  title        = {The {DRAT} format and DRAT-trim checker},
  journal      = {CoRR},
  volume       = {abs/1610.06229},
  year         = {2016}
}

@inproceedings{CzernerEK24,
  author       = {Philipp Czerner and
                  Javier Esparza and
                  Valentin Krasotin},
  editor       = {Naoki Kobayashi and
                  James Worrell},
  title        = {A Resolution-Based Interactive Proof System for {UNSAT}},
  booktitle    = {Foundations of Software Science and Computation Structures - 27th
                  International Conference, FoSSaCS 2024, Held as Part of the European
                  Joint Conferences on Theory and Practice of Software, {ETAPS} 2024,
                  Luxembourg City, Luxembourg, April 6-11, 2024, Proceedings, Part {II}},
  series       = {Lecture Notes in Computer Science},
  volume       = {14575},
  pages        = {116--136},
  publisher    = {Springer},
  year         = {2024},
  doi          = {10.1007/978-3-031-57231-9_6},
  bibsource    = {dblp computer science bibliography, https://dblp.org}
}

@inproceedings{Cruz-FilipeHHKS17,
  author       = {Lu{\'{\i}}s Cruz{-}Filipe and
                  Marijn J. H. Heule and
                  Warren A. Hunt Jr. and
                  Matt Kaufmann and
                  Peter Schneider{-}Kamp},
  editor       = {Leonardo de Moura},
  title        = {Efficient Certified {RAT} Verification},
  booktitle    = {Automated Deduction - {CADE} 26 - 26th International Conference on
                  Automated Deduction, Gothenburg, Sweden, August 6-11, 2017, Proceedings},
  series       = {Lecture Notes in Computer Science},
  volume       = {10395},
  pages        = {220--236},
  publisher    = {Springer},
  year         = {2017},
  noUrlBecauseOfDoi          = {https://doi.org/10.1007/978-3-319-63046-5\_14},
  doi          = {10.1007/978-3-319-63046-5\_14},
  bibsource    = {dblp computer science bibliography, https://dblp.org}
}

@book{AB09,
  author       = {Sanjeev Arora and
                  Boaz Barak},
  title        = {Computational Complexity - {A} Modern Approach},
  publisher    = {Cambridge University Press},
  year         = {2009}
}
